\documentclass[aps,pre,twocolumn,superscriptaddress,showpacs,floatfix,nofootinbib]{revtex4-1}
\usepackage{dcolumn}
\usepackage{amsmath}
\usepackage{amssymb}
\usepackage{amsthm}
\usepackage{graphicx}
\usepackage{bm}
\usepackage[T1]{fontenc}
\usepackage{color}
\usepackage{url}
\usepackage[bf]{subfigure}
\usepackage{rotating}
\usepackage{multirow}

\usepackage{scalefnt}
\usepackage{times}
\usepackage{mathtools}
\usepackage{bbm}

\usepackage{float}
\usepackage{enumerate}
\usepackage[makeroom]{cancel}

\usepackage{tikz}
\usetikzlibrary{topaths,calc}

\usepackage[bookmarks, breaklinks]{hyperref} 

\theoremstyle{definition}

\newcommand{\A}{\mathbf{A}}

\newcommand{\D}{\mathbf{D}}

\newcommand{\Q}{\mathcal{A}}
\newcommand{\Dq}{\mathcal{D}}
\newcommand{\Lq}{\mathcal{L}}
\newcommand{\Pq}{\mathcal{P}}

\newcommand{\Prob}[1]{\mathbb{P}\left( #1 \right)}
\newcommand{\probP}[1]{\text{I\kern-0.15em P} \left( #1 \right)}
\newcommand{\bigzero}{\mbox{\normalfont\Large\bfseries 0}}

\begin{document}

\title{From unbiased to maximal entropy random walks on hypergraphs}

\author{Pietro Traversa}
\affiliation{Institute for Biocomputation and Physics of Complex Systems (BIFI), University of Zaragoza, 50018 Zaragoza, Spain}
\affiliation{Department of Theoretical Physics, University of Zaragoza, 50018 Zaragoza, Spain}
\affiliation{CENTAI Institute, Turin, Italy}

\author{Guilherme Ferraz de Arruda}
\affiliation{CENTAI Institute, Turin, Italy}

\author{Yamir Moreno}
\affiliation{Institute for Biocomputation and Physics of Complex Systems (BIFI), University of Zaragoza, 50018 Zaragoza, Spain}
\affiliation{Department of Theoretical Physics, University of Zaragoza, 50018 Zaragoza, Spain}
\affiliation{CENTAI Institute, Turin, Italy}

\begin{abstract}

    Random walks have been intensively studied on regular and complex networks, which are used to represent pairwise interactions. Nonetheless, recent works have demonstrated that many real-world processes are better captured by higher-order relationships, which are naturally represented by hypergraphs. Here, we study random walks on hypergraphs. Due to the higher-order nature of these mathematical objects, one can define more than one type of walks. In particular, we study the unbiased and the maximal entropy random walk on hypergraphs, emphasizing their similarities and differences. We characterize these processes through their stationary distributions and the associated hitting times, which we obtain analytically. To illustrate our findings, we present a toy example and conduct extensive analyses of artificial and real hypergraphs, providing insights into both their structural and dynamical properties. We hope that our findings motivate further research extending the analysis to different classes of random walks as well as to practical applications. 
    
\end{abstract}

\maketitle

\section{Introduction}

One of the main frameworks used to study and describe complex systems is network theory, which has been greatly developed during the last two decades. Despite this development and its success to represent and understand a plethora of real systems, most network methods are constrained to systems with pairwise interactions. Recently, attention was raised to higher-order interactions, arguing that rich data are revealing more complex relationships among nodes that may not be captured by models based on pairwise interactions~\cite{battiston_physics_2021,Lambiotte2019}. This claim has been supported through a series of works. In~\cite{Mulas2020, Arruda2020b}, through linear stability analysis, hypergraphs' stability was evaluated, emphasizing some of the key differences and similarities between graphs and hypergraphs.  From a modeling point of view, the need to consider higher-order interactions has also been recently reinforced by theoretical approaches involving phenomena such as social contagion~\cite{Bodo2016, Iacopini2019, Arruda2020}, evolutionary game dynamics~\cite{Alvarez-Rodriguez2020} synchronization~\cite{Ginestra2020, Maxime2020, Gallo2020} and random walks~\cite{Zhou2006, Chitra2019, Carletti2020, Hayashi2020}, the latter being the main focus of this contribution. 

The study of hypergraphs is also important in other fields of research beyond physics or mathematics. For instance, in the area of machine learning research, hypergraphs have been used in classification, clustering, and embedding techniques~\cite{Zhou2006, Hayashi2020}. Moreover, a hypergraph convolutional neural network (HGCN) has been proposed~\cite{Feng2019, Yadati2019}. Of relevance for the present work, the authors of \cite{Chitra2019} hypothesized that machine learning algorithms could benefit from further studies of hypergraphs and, more specifically, on random walks. The aforementioned list of works is not exhaustive but it shows the increasing interest that data-rich and higher-order approaches are attracting. Nonetheless, despite this interest, the study of higher-order systems is arguably in its infancy. Thus, it is of utmost importance to build most of the theoretical tools that will allow us to study and develop more complex and realistic processes in the near future. In this context, a random walk process is simple enough to provide new insights and results while capturing this type of system's higher-order nature. 

Random walks are paradigmatic, being interesting both from theoretical and practical points of view. They are probably the most fundamental stochastic processs~\cite{Masuda2017}, serving as a model for a variety of phenomena, including diffusion, social interactions, opinions~\cite{Masuda2017} and providing handy insights that can be used in many different contexts. In network theory, this process and its variants are reasonably well studied~\cite{Burda2009, Zhongzhi2012, Lin2014, Guo2016, Masuda2017, Zhongzhi2018}. However, in hypergraphs, this process just got recent attention due to its applications in machine learning and physics. As Carletti et al. mentioned in~\cite{Carletti2020}, probably the first random walk defined on hypergraphs was proposed by Zhou et al. in~\cite{Zhou2006}. In this paper, the authors were concerned about using such a process in machine learning techniques such as clustering, classification, and embedding. 
In~\cite{Hayashi2020}, also focusing on machine learning applications, Koby Hayashi et al. proposed a clustering framework using hypergraph-structured data-based and edge-dependent vertex weights random walks. Here, we are interested in the physical aspects and insights that random walks can bring to the analysis of hypergraphs' structure and dynamics. Thus, the most straightforward choices are the edge-independent vertex weights with trivial weights. In this configuration, the fundamental structural aspects will play a significant role. We note that in ~\cite{Carletti2020}, the authors followed a similar approach. 

Here we focus on two classes of random walks, the unbiased random walk (URW), where only local information is used by the walker, and the maximal entropy random walk (MERW), where global information is used to construct the transition probabilities. Formally, the walker in an unbiased random walk~\footnote{In the literature, this class of random walks is also called a general random walk (GRW).} makes a succession of uniformly random decisions using only local information (the node degree). On the other hand, in the MERW, the walker uniformly chooses a path that maximizes the entropy among all the possible paths of fixed length~\cite{Burda2009, Duda2012}. The construction of such a process requires complete knowledge about the structure, here expressed by the leading eigenvector of the adjacency matrix. Comparatively, the first is a local process, while the second is a non-local dynamics. In the context of network analysis, MERW was applied to the analysis of networks with limited information~\cite{Sinatra2011}, community detection~\cite{Ochab2013}, link predictions~\cite{Li2011}, and on the definition of centrality measures~\cite{Delvenne2011}.

Our main contribution is the generalization of maximal entropy random walks to hypergraphs, revealing different behaviors. Another key contribution is the extensions of network results for random walks on hypergraphs and their analysis in artificial and real scenarios. We establish our derivations from the observation that random walks on complex networks are equivalent to the same processes on hypergraphs, up to some small details and constraints~\cite {Chitra2019, Hayashi2020}. Similar observations were also made in~\cite{Carletti2020}. Thus, our paper complements and extends the findings of Refs.~\cite{Burda2009, Chitra2019, Carletti2020, Hayashi2020}. Our construction of the random walk provides a different interpretation of the process if compared to~\cite{Carletti2020}, as we explicitly define the step as a sequence of two processes, similar to Ref.~\cite{Chitra2019, Hayashi2020}. Formally, taking advantage of network theory results, we extended the maximal entropy random walk and focused on comparing this random walk and the unbiased version. Moreover, we concentrate our efforts on the analysis of stationary distributions and hitting times. Another contribution of our work is the numerical experiments, in which we provide a series of examples, ranging from a small toy example, emphasizing key peculiarities of random walks in hypergraphs, to artificial and real cases. For the artificial experiments, we remark that we were able to evaluate heterogeneity in both the cardinality distribution and the degree distribution.

Our paper is divided as follows. In the next section, we define hypergraphs and discuss their representations. Next, in Sec.~\ref{sec:RW}, we define the random walks, focusing on the unbiased random walks, in sub-section~\ref{sec:URW}, and the maximal entropy, in sub-section~\ref{sec:MERW}. Complementary, in sub-section~\ref{sec:uniform}, we discuss the particularities of uniform and regular structures, while in Sec.~\ref{sec:numerical}, we numerically evaluate synthetic and real hypergraphs. Specifically, in  sub-section~\ref{sec:toy}, we present a simple toy example that allows us to discuss the main differences among the classes of random walks studied here and in sub-section~\ref{sec:models} we describe the model we used to generate the artificial hypergraphs. To sum up, in Sec.~\ref{sec:analysis}, we present a short discussion about our findings and their relation with the literature, followed by the conclusions.

\section{Hypergraphs: definitions and representation}\label{Sec: representation}

A hypergraph, $\mathcal{H} = \{ \mathcal{V}, \mathcal{E}\}$, is a mathematical structure that extends the concept of a graph. It is composed of a set of nodes, $\mathcal{V} = \{ v_i \}$, with $v_i \in \mathbb{Z}^+$, and a multiset of hyperedges $\mathcal{E} = \{ e_j \}$, where $e_j$ is a non-empty subset of $\mathcal{V}$ with arbitrary cardinality $|e_j|$. The number of nodes in the hypergraph is denoted by $N = |\mathcal{V}|$, and the maximum cardinality of the hyperedges is given by $e_{\max} = \max \left( |e_j| \right)$.
 We also denote $\mathcal{E}_i$ as the multiset of hyperedges that contain the node $i$. A hypergraph is considered to be simple if there are no repeated hyperedges, i.e. if $\mathcal{E}$ is a set rather than a multiset. If $e_{\max} = 2$, the hypergraph reduces to a standard graph, whereas one recovers a simplicial complex if, for each hyperedge with $|e_j| > 2$, its subsets are also contained in $\mathcal{E}$.
 The degree of node $i$, $k_i$, is defined as the number of hyperedges that contain this node. Conversely, the number of neighbors of node $i$, $n_i$ is defined as the number of unique nodes that share a hyperedge with $i$. We remark that these two concepts coincide in graphs, but they might be different in hypergraphs.

A hypergraph $\mathcal{H}$ can be represented by the \emph{incidence matrix}, $\mathcal{I} \in \mathbb{R}^{N \times M}$, which is defined as
\begin{equation}
    \mathcal{I}_{ij} =
    \begin{cases}
        1 \hspace{5mm} \text{if} \hspace{2mm} v_i \in e_j \\
        0 \hspace{5mm} \text{if} \hspace{2mm} v_i \notin e_j
    \end{cases}.
\end{equation}
From this representation, we can define the \emph{counting adjacency matrix}~\cite{Carletti2020, Battiston2020}, $\A^{count} \in \mathbb{R}^{N \times N}$, given as
\begin{equation} \label{eq:adjacency_count}
 \A^{count} = \mathcal{I} \mathcal{I}^T - \D ,
\end{equation}
where $\D = \text{diag} \left( k_i \right)$. Each element $\A^{count}_{ij}$ is the number of hyperedges shared by nodes $i$ and $j$.
The \emph{normalized adjacency matrix}~\cite{Banerjee2020}, $\A^{norm} \in \mathbb{R}^{N \times N}$, is defined as 
\begin{equation} \label{eq:adjacency_norm}
    \A_{ik}^{norm} = \sum_{\substack{e_j \in \mathcal{E} \\ v_i, v_k \in e_j \\ v_i \neq v_k }} \frac{1}{|e_j| - 1} .
\end{equation}

In this formulation, the degree of node $i$ is computed as $k_i = \sum_{k=1}^N \A_{ik}^{norm}$. We remark that in the original definition in~\cite{Banerjee2020}, this matrix was not referred to as the normalized adjacency matrix. However, to emphasize its differences with respect to the counting adjacency matrix,  we have referred to it as such here. Both matrices can be considered as projections of the hypergraph onto a graph, where hyperedges are represented as cliques. The information about the hyperedges is retained in the edge weights of the projected graph, although the counting and normalized projections assign different meanings to these weights.

\section{Random walks}
\label{sec:RW}

A \emph{walk}~\cite{Banerjee2020} $v_{i_0}-v_{i_l}$ of length $l$ between two vertices $v_{i_0},v_{i_l}\in \mathcal{V}$ in a hypergraph $\mathcal{H}$ is an alternating sequence $v_{i_0} e_{j_1}  v_{i_1} e_{j_2} \dots v_{i_{l-1}} e_{j_l} v_{i_l}$ of distinct pairs of vertices and hyperedges, such that $v_{i_{k-1}},v_{i_k} \in e_{j_k}$ for $k=i,\dots,l$.  A \emph{step} is a walk of length one.  A \emph{random walk} (RW) is a stochastic process, which describes a walk consisting of a succession of random steps.
Here, we focus on \emph{Markovian random walks}, where the next step is dependent only on the current state of the process. At each time step $t$, the random walk process proceeds as follows:
\begin{itemize}
    \item pick an edge $e \in \mathcal{E}_{i_t}$ with some probability $p_{v_{i_t}}(e)$,
    \item pick a vertex $v \in e $ with some probability $p_e(v)$,
    \item move to $v_{i_{t+1}}=v$ at time $t+1$.
\end{itemize}
The probabilities associated with choosing an edge and a vertex may vary depending on the type of random walk considered.
For instance, the walker at node $v_{i_t}$ first chooses a hyperedge $e$, then, inside this hyperedge, it chooses uniformly a different node $v_{i_{t+1}} \in e \setminus \{v_{i_t}\}$. This process is related to the graph-projection given by Eq.~\eqref{eq:adjacency_norm}. We denote this as the higher-order step and the generated process as a higher-order random walk (HORW). This type of process was initially studied in~\cite{Chitra2019, Hayashi2020, Banerjee2020} and also explored as an unbiased random walk. 
Another possible choice is to consider the next step probability for a walker at node $v_{i_t}$ to be proportional to the number of hyperedges between $v_{i_t}$ and $v_{i_{t+1}}$. We call this the projected step and the generated process as a projected random walk (PRW). In this case, the walk takes place in the graph-projection defined by Eq.~\eqref{eq:adjacency_count}. This type of higher-order step was initially explored in~\cite{Carletti2020} in the form of unbiased random walks. We stress that this is not a two-event process, as we have no information on which hyperedge the walk is moving through, but only on how many hyperedges two nodes share. As a consequence, this type of higher-order step may not be sensitive to some higher-order structures. Note that, in pairwise relations, both formulations fall into the standard definitions of random walks in graphs.

\begin{table}[t!] \scalefont{0.92}
    \caption{Definition of the fundamental matrices used for the two types of random walks considered: the projected random walk (P-RW) and the higher-order random walk (HO-RW). The probability transition matrices for the unbiased and maximal entropy cases are analyzed in detail in sections~\ref{sec:URW} and~\ref{sec:MERW} respectively.}
    \begin{tabular}{|c|c|c|}
        \hline
        Type & \textbf{P-RW} & \textbf{HO-RW}  \\
        \hline
        Adjacency & $\Q = \A^{count}$ & $\Q = \A^{norm}$  \\
        \hline
        Diagonal & $\Dq = \text{diag}\left( \sum_{k=1}^N \A^{count}_{ik} \right)$ & $\Dq = \text{diag}\left( \sum_{k=1}^N \A_{ik}^{norm} \right)$\\
        \hline
        Laplacian & $\Lq = \Dq - \A^{count}$ & $\Lq = \Dq - \A^{norm}$ \\
        \hline
        \begin{minipage}{1.4cm}
            Probability transition (URW)
        \end{minipage} & \multicolumn{2}{c|}{$\Pq^{URW} = \Dq^{-1} \Q$}\\
        \hline
        \begin{minipage}{1.4cm}
            Probability transition (MERW)
        \end{minipage} &  \multicolumn{2}{c|}{$\Pq_{ij}^{MERW} = \frac{\Q_{ij}}{\lambda} \frac{\psi_j}{\psi_i}$}\\
        \hline
    \end{tabular}
    \label{tab:definitions}
\end{table}

As we defined a random walk where the nodes are the states, all our quantities of interest can be derived in terms of $N \times N$ matrices. This argument was formally provided in Ref.~\cite{Chitra2019}, Theorem 16, where, by using the time reversibility property of Markov Chains, the authors proved that the non-lazy \footnote{The lazy random walk allows the walker to stay at the current node, while the non-lazy random walk does not allow it. In other words, $\Pq_{ii} \geq 0$ for the lazy and $\Pq_{ii} = 0$ for the non-lazy.} random walk on a hypergraph is equivalent to the non-lazy random walk on a graph, provided that there are trivial vertex weights. Hence, using the mapping between random walks in hypergraphs and graphs (under the constraints mentioned above), the theory of random walks in weighted networks can be reinterpreted in our context. In the following sections, we define two types of random walks: unbiased and maximal entropy random walks. We can also use the two types of steps previously discussed, the projected and higher-order steps for each class of random walks. Thus, to make our notation lighter, we define the adjacency, $\Q$, the Laplacian, $\Lq$, and the probability transition matrix, $\Pq$ accordingly. Their definitions are given in Table~\ref{tab:definitions}. We also use superscripts to denote the type of step and random walks when necessary.

\subsection{Unbiased random walks}
\label{sec:URW}

For the sake of generality and comparison, we analyze the unbiased random walk (URW), complementing some literature results. In this process, given a step definition, there is no bias towards a given direction. From the Markovian formulation, the \emph{stationary distribution} is expressed as
\begin{equation}
    \pi^T = \pi^T \Pq^{URW},
\end{equation}
where $\Pq^{URW} = \Dq^{-1} \Q$ is the probability transition matrix, $\Dq = \text{diag} \left( \sum_{j=1}^N \Q_{ij} \right)$ is a diagonal matrix, and $\pi$ is the normalized eigenvector associated to the leading eigenvalue, $\sum_{i=1}\pi_i = 1$. Explicitly, this distribution is given as
\begin{equation}
    \label{eq:pi}
    \pi_i = \frac{\sum_{j=1}^N \Q_{ij}}{\sum_{\substack{i=1 \\ j=1}}^N \Q_{ij}}.
\end{equation}

Aside from the stationary distribution, other quantities of interest are the hitting times. We denote by $0=\sigma_1<\sigma_2\leq \cdots \leq \sigma_N$ the $N$ eigenvalues of the Laplacian matrix, $\Lq$, and by $\mu_1, \mu_2, \cdots, \mu_N$ their corresponding normalized eigenvectors, whose components are $\mu_i=(\mu_{i1}, \mu_{i2}, \cdots, \mu_{iN})^{\top}$ for $i=1,2, \cdots, N$. For a random walker starting from node $v_i$, the expected time to hit node $v_j$ is expressed as~\cite{Zhongzhi2012, Zhongzhi2018}
\begin{equation}\label{eq:Tij_unbiased}
T_{ij}^{URW} = \sum_{z=1}^{N} k_z\sum_{k=2}^{N}\frac{1}{\sigma_k}(\mu_{ki}\mu_{kz}-\mu_{ki}\mu_{kj}-\mu_{kj}\mu_{kz}+\mu_{kj}^2).
\end{equation}
Complementary, assuming that the node $v_j$ is the trap node, the partial mean hitting time is given as
\begin{equation}\label{eq:Tj_unbiased}
T_{j}^{URW} = \frac{N}{N-1}\sum_{k=2}^{N}\frac{1}{\sigma_k}\left(2E\times\mu_{kj}^2-\mu_{kj}\sum_{z=1}^{N}k_z\mu_{kz}\right),
\end{equation}
where $E = \sum_{\substack{i=1 \\ j=1}}^N \Q_{ij}$. 
Finally, the global mean first passage time can be obtained as
\begin{equation}\label{eq:T_unbiased}
\langle T^{URW} \rangle = \frac{2E}{N-1}\sum_{k=2}^{N}\frac{1}{\sigma_k}.
\end{equation}
Interestingly, the hitting times are fully characterized using only the spectral properties of the Laplacian matrix, while the stationary distribution depends only on the probability transition matrix.

\subsection{Maximal entropy random walk}
\label{sec:MERW}

Another interesting case is the maximal entropy random walk, where the walker is biased towards the direction that maximizes the entropy of possible trajectories.

In this case, the probability transition matrix is defined as~\cite{Burda2009}
\begin{equation}\label{eq: MERW}
    \Pq_{ij}^{MERW} = \frac{\Q_{ij}}{\lambda_1} \frac{\psi_{1j}}{\psi_{1i}}.
\end{equation}
We denote by $\lambda_1 \geq \lambda_2 \geq \cdots \geq \lambda_N$ the eigenvalues of $\Q$ and by $\psi_1 \geq \psi_2 \geq \cdots \geq \psi_N$ the associated eigenvectors, whose normalized components are $\psi_i=(\psi_{i1}, \psi_{i2}, \cdots, \psi_{iN})^{\top}$ for $i=1,2, \cdots, N$. We obtain the stationary distribution as
\begin{equation}
    \label{eq:phi}
    \phi_i = \psi_{1i}^2,
\end{equation}
where the normalization $\sum_{i=1}\psi_{1i}^2 = 1$ must hold.

Mathematically, the expected time to hit $v_j$, starting from $v_i$ is given as
\begin{equation}\label{eq:Tij_ME}
T_{ij}^{MERW}=\frac{1}{\psi_{1j}^2}\sum_{k=2}^{N}\frac{\lambda_1}{\lambda_1-\lambda_k}\left(\psi_{kj}^2-\psi_{ki}\psi_{kj}\frac{\psi_{1j}}{\psi_{1i}}\right).
\end{equation}
The partial mean hitting time to reach $j$ is
\begin{align}\label{eq:Tj_ME}
T_j&^{MERW}= \nonumber \\
&\frac{1}{\psi_{1j}^2(N-1)}\sum_{k=2}^N\frac{\lambda_1}{\lambda_1-\lambda_k}\left(N\psi_{kj}^2-\psi_{kj}\psi_{1j}\sum_{i=1}^N\frac{\psi_{ki}}{\psi_{1i}}\right),
\end{align}
and, the global mean hitting time is
\begin{align}\label{eq:T_ME}
\langle T^{MERW} \rangle&=\frac{1}{N(N-1)}\sum_{j=1}^N\frac{1}{\psi_{1j}^2} \times \nonumber\\
&\quad\sum_{k=2}^N\frac{\lambda_1}{\lambda_1-\lambda_k}\left(N\psi_{kj}^2-\psi_{kj}\psi_{1j}\sum_{i=1}^N\frac{\psi_{ki}}{\psi_{1i}}\right).
\end{align}
In contrast with the unbiased case, in the maximal entropy random walk, both the hitting times and the stationary distribution depend only on the eigenvalues and eigenvectors of adjacency matrix $\Q$~\cite{Lin2014, Zhongzhi2018}.

\subsection{Uniform and regular hypergraphs}
\label{sec:uniform}

In \emph{uniform hypergraphs}, all the hyperedges have the same cardinality, i.e. $|e_j| = c$ for all $j \in \{1, 2, \cdots M\}$. From the spectral viewpoint, the spectra of both the counting and the normalized adjacency matrices are the same, with a difference of only a scaling factor $(c - 1)$. Similar arguments can also be applied to the Laplacian and the probability transition matrices. Consequently, for a given class of random walks, both the projected and the higher-order steps have the same stationary distributions and hitting times. Although uniform hypergraphs are relatively simpler structures, it does not imply that they are trivial. Even without heterogeneity in the distribution of cardinalities, the degree distribution can still be heterogeneous. We numerically explored this case in Sec.~\ref{sec:numerical_res}.

Furthermore, the MERW and the URW are equivalent in uniform and regular hypergraphs, where all the nodes in $\mathcal{H}$ have the same degree. In this case, the spectra of the Laplacian matrix can be described by the spectra of the adjacency matrix up to a scale and translation, thus, implying that both classes of random walks present the same behavior. Finally, it is worth mentioning that, in~\cite{Cooper2011}, the authors formally derived the cover times for a class of regular and uniform hypergraphs, providing exact expression as well as asymptotic results.

\section{Numerical experiments}
\label{sec:numerical}

In this section, we complement our analysis with numerical experiments. First, in Section~\ref{sec:artificial}, we focus on artificial hypergraphs, evaluating both the distribution of cardinalities and the degree distribution, where we consider uniform hypergraphs. Next, in section~\ref{sec:real} we show an example of a real hypergraph, extending our analysis to cases where non-trivial correlations are present.

\subsection{Artificial hypergraphs}
\label{sec:artificial}

First, we present a toy example that allows us to comment on both the differences between the step definitions and the random walks. Next, in sub-sections~\ref{sec:models} and~\ref{sec:numerical_res}, we describe and evaluate a series of synthetic hypergraphs with different levels of heterogeneity.

\subsubsection{A toy example}
\label{sec:toy}

\begin{figure}[t!]
    \includegraphics[width=\linewidth]{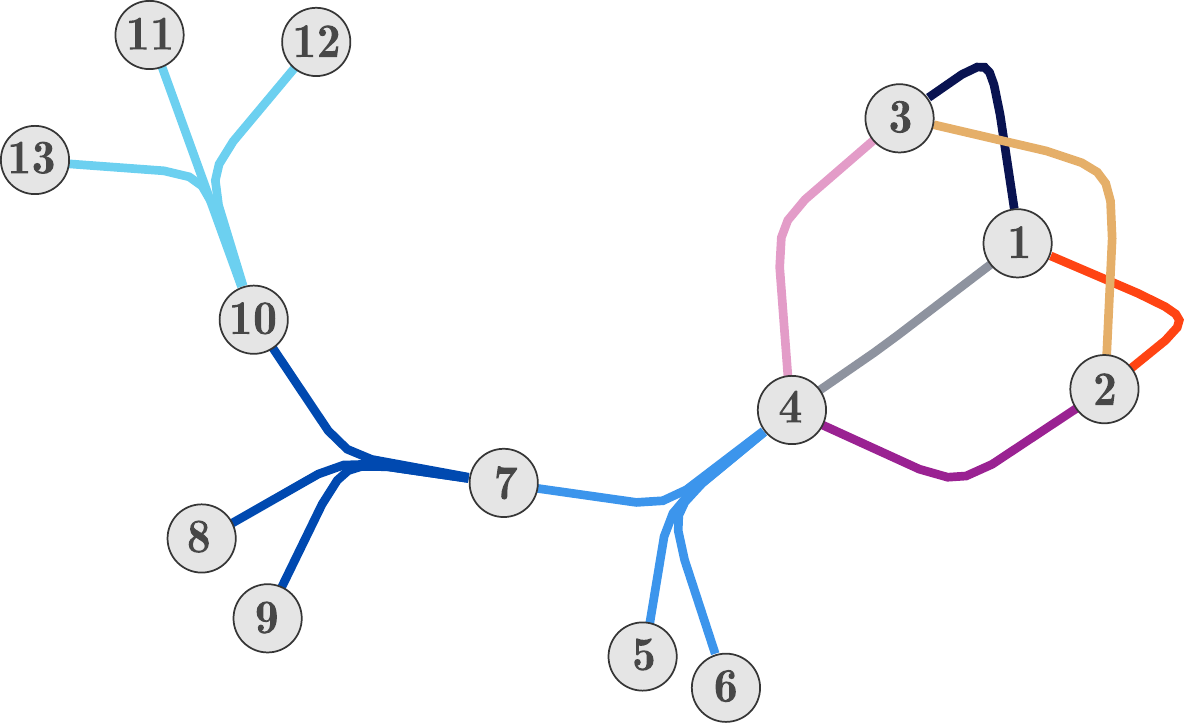}
    \caption{A toy hypergraph, $\mathcal{H} = \{ \mathcal{V}, \mathcal{E}\}$, with $N=13$ nodes, $M = 9$ hyperedges, where $\mathcal{E} = \{e_1, e_2, ..., e_9 \}$ and $e_1 = \{1, 2\}$, $e_2 = \{1, 3\}$, $e_3 = \{1, 4\}$, $e_4 = \{2, 3\}$, $e_5 = \{2, 4\}$, $e_6 = \{3, 4\}$, $e_7 = \{4, 5, 6, 7\}$, $e_8 = \{7, 8, 9, 10\}$, $e_9 = \{10, 11, 12, 13\}$. The hyperedges are color coded.}
    \label{fig:Toy}
\end{figure}

\begin{figure}[t!]
    \includegraphics[width=\linewidth]{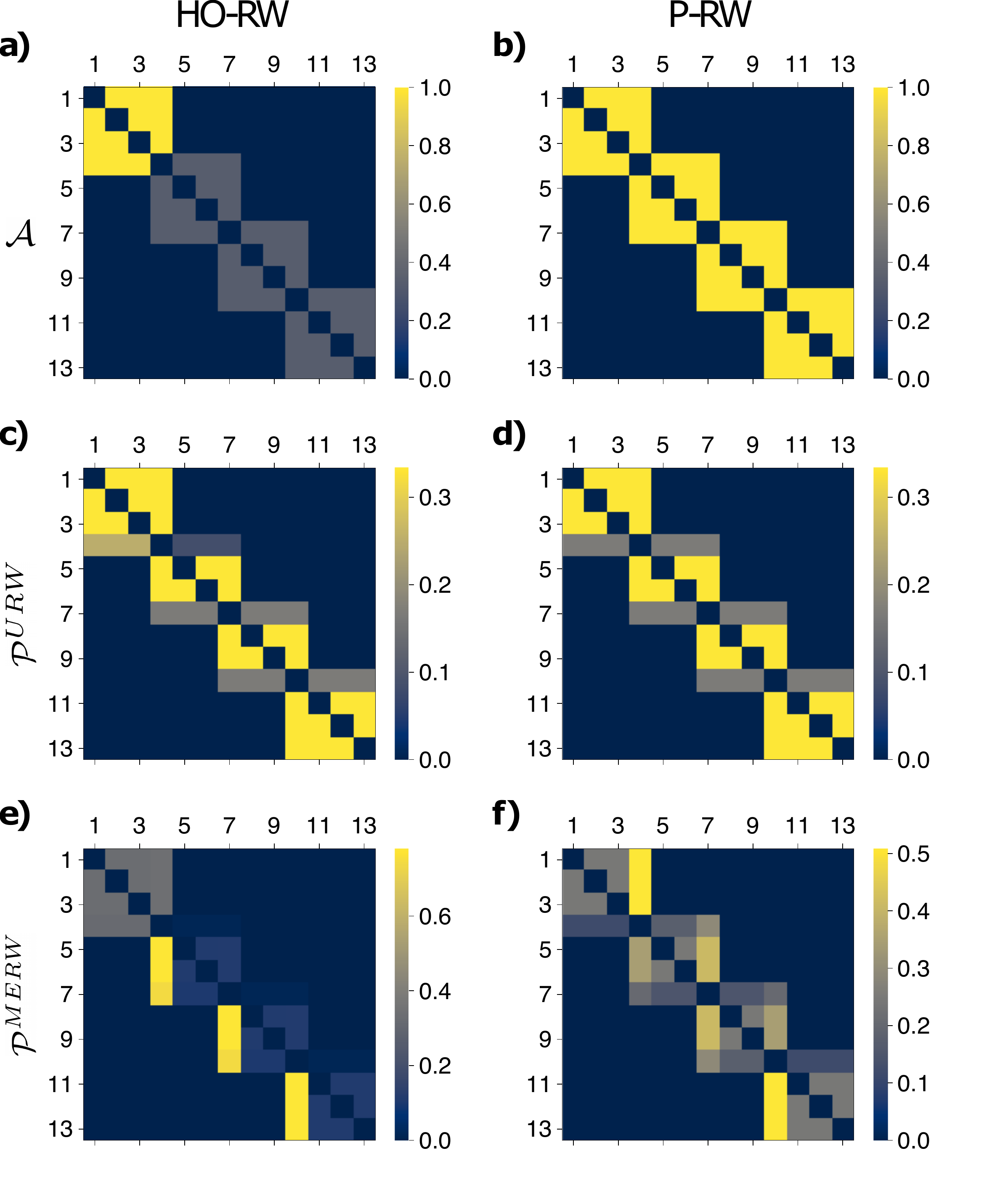}
    \caption{Graphical representation of the adjacency and probability transition matrices for the toy's  example. In (a), (c), and (e) we show the matrices related to the higher-order step, while in (b), (d), and (f) the projected step. In (a) and (b) the adjacency matrices, $\A^{norm}$ and $\A^{count}$ respectively. In (c) and (d) the unbiased random walk probability transition matrix, $\Pq^{URW}$, while in (e) and (f) the maximum entropy random walk transition matrix, $\Pq^{MERW}$.}
    \label{fig:Toy_A_P}
\end{figure}

We analyze a small toy example of a hypergraph to highlight the differences among the random walks studied here. We consider a hypergraph with $N = 13$ nodes and $M = 9$ hyperedges, as depicted in Fig.~\ref{fig:Toy}. This hypergraph is ``nearly symmetric'' with respect to node $7$, as the structure $\mathcal{E}^{\boxtimes} = \{e_1, e_2, e_3, e_4, e_5, e_6 \}$ is a projected clique of a hyperedge with cardinality $4$ and composed by nodes $1$, $2$, $3$ and $4$. 

Figure~\ref{fig:Toy_A_P} shows the adjacency and probability transition matrices.
We observe that the counting adjacency matrix in Fig.~\ref{fig:Toy_A_P} (b) does not distinguish between the structures $e_9 = \{10, 11, 12, 13\}$ and the set of hyperedges $\mathcal{E}^{\boxtimes}$. On the other hand, the representation given by Eq.~\eqref{eq:adjacency_norm}, in Fig.~\ref{fig:Toy_A_P} (a), weights these two types of structures differently. 
Regarding the transition matrices $\Pq^{URW}$ in Fig.~\ref{fig:Toy_A_P} (c) and (d),  the differences are relatively small, with node $4$ playing a slightly different role. However, for the $\Pq^{MERW}$ in Fig.~\ref{fig:Toy_A_P} (e) and (f), the differences are more pronounced. Specifically, in the higher-order case, the walker is more likely to remain within the set of hyperedges $\mathcal{E}^{\boxtimes}$, in contrast to the projected step.

\begin{figure}[t!]
    \includegraphics[width=\linewidth]{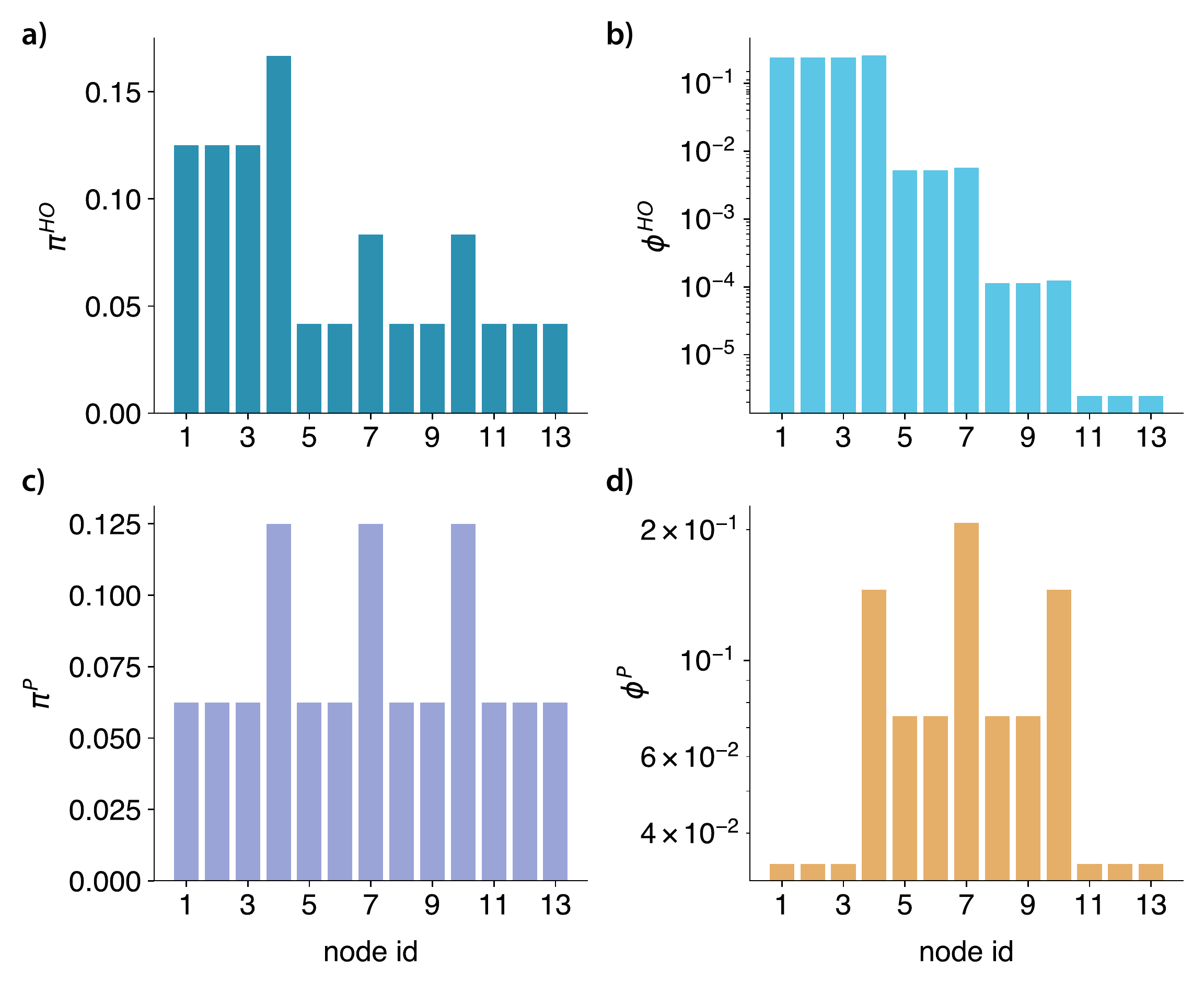}
    \caption{Toy's  example stationary distribution for the unbiased random walk with the higher-order step in (a), the maximal entropy with the same step in (b), the unbiased random walk with projected step in (c), and the maximal entropy random walk with the same step in (d).}
    \label{fig:Toy_Densities}
\end{figure}

In Figure~\ref{fig:Toy_Densities}, we examine the stationary distributions of the four types of random walks considered. Figure~\ref{fig:Toy_Densities} (a) and (b) clearly show the effects of asymmetry in the higher-order step cases.  In contrast, the projected step creates a symmetry that is reflected in the stationary distributions, which can be seen in Figure~\ref{fig:Toy_Densities} (c) and (d). This effect is evident in the probabilities of nodes 4, 7, and 10 in both the unbiased and maximum entropy random walks. 

\begin{figure}[t!]
    \includegraphics[width=\linewidth]{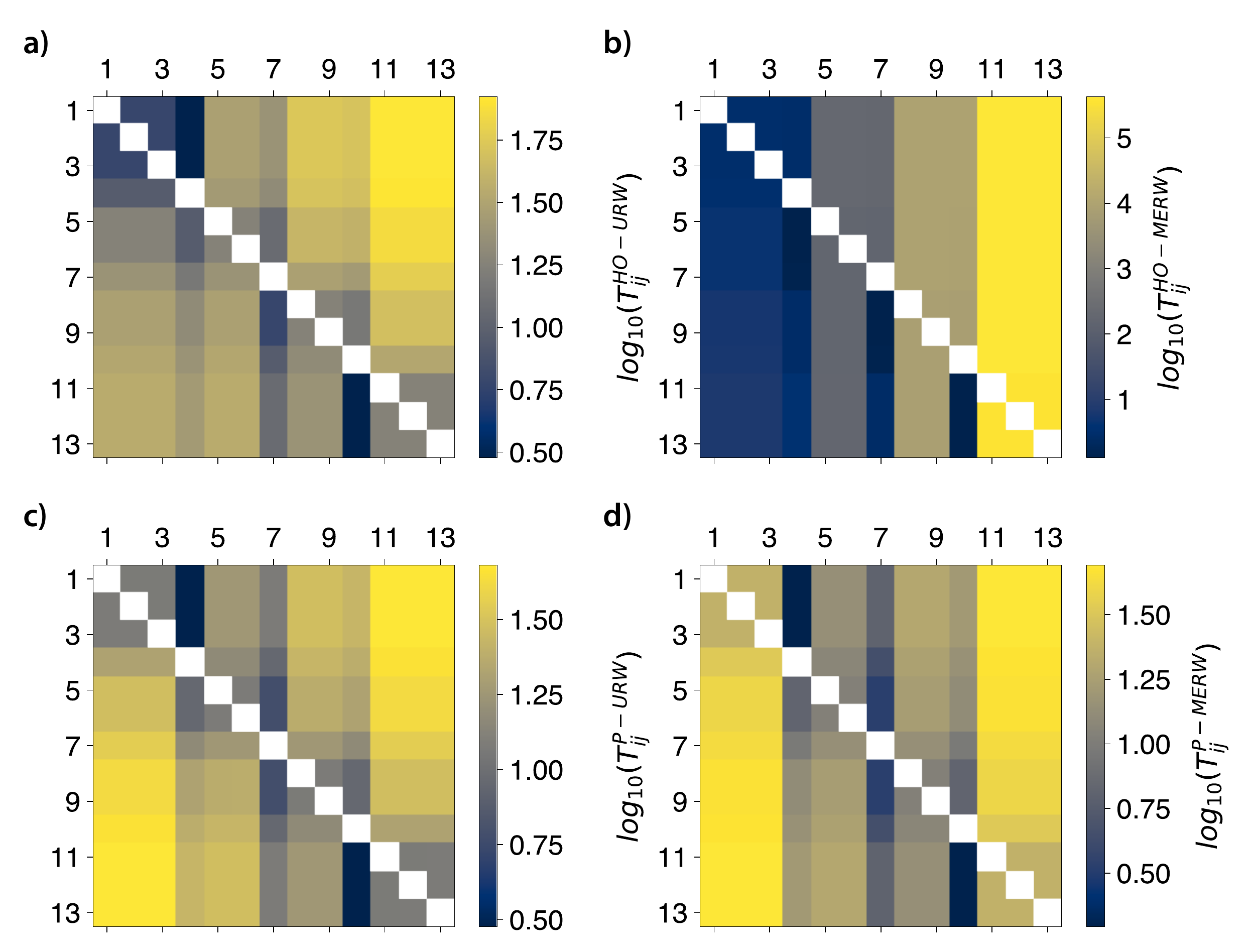}
    \caption{Toy's  example hitting times, $T_{ij}$, for the different types of random walk studied. In (a) and (c) for the unbiased random walk and in (b) and (d) for the maximum entropy random walk. Complementary, in (a) and (b) the higher-order step, while in (c) and (d) the projected step.}
    \label{fig:Toy_Tij}
\end{figure}

Complementary, the expected time to hit $v_j$, starting from $v_i$ is represented in Fig.~\ref{fig:Toy_Tij}. In the projected case, Fig.\ref{fig:Toy_Tij} (c) and (d), both types of random walks exhibit the same behavior. Additionally, it is evident that the two types of steps are slightly different, where the higher-order step creates a bias towards the subset $\mathcal{E}^{\boxtimes}$. This effect is even more apparent in the maximum entropy case, where the time to reach the nodes $1$, $2$, $3$, and $4$ is considerably shorter than that for nodes ${10}$, ${11}$, ${12}$, and ${13}$. Moreover, we remark on the particular role of nodes $4$, $7$, and ${10}$, which serve as bridges. Thus, the probabilities of getting to these nodes typically present a strong dependency on the origin node.

\subsubsection{Hypergraph models}
\label{sec:models}

We begin by considering the random uncorrelated model, where we generate hypergraphs with an arbitrary distribution of cardinalities without controlling the degree distribution. Specifically, we fix the number of nodes, $N$, and hyperedges, $M$, and sample their cardinalities from a given distribution, $P(|e_j|)$. To generate a hyperedge $e_j$, we uniformly sample $|e_j|$ nodes from $\mathcal{V}$.
This procedure results in a random hypergraph, which may have multiple connected components. To ensure a single connected component, we use a brute force algorithm that tries $K = 200$ times to find a connected component~\footnote{A connected component in a hypergraph can be obtained by using the standard graph algorithms in the count or normalized adjacency matrices.} with $N$ nodes. If all $K$ trials are unsuccessful, it increases or decreases the number of nodes by $n^* = 10$ and repeats the procedure $K$ times. The algorithm stops once the first connected component is a hypergraph with $N$ nodes. Although this method can introduce small fluctuations in the number of hyperedges, it maintains the number of nodes fixed. 

Using the above-described model, we constructed both Poisson, $P(|e_j|) \sim \text{Poisson}(\beta)$, and power-law (PL) distributions, $P(|e_j|) \sim |e_j|^{-\gamma}$. In terms of cardinalities, the Poisson distribution generates more homogeneous hypergraphs, where we can control the average cardinality. On the other hand, the PL distribution represents a class of heterogeneous hypergraphs, whose heterogeneity can be controlled by the parameter $\gamma$. While more sophisticated methods, as proposed in~\cite{Chodrow2020}, may be available, they can incur higher computational costs, making them impractical for large sample sizes. Here, for each experiment, we considered $n_{\text{runs}} = 10^3$ independently generated hypergraphs.

In addition to the heterogeneity of cardinalities, we also consider uniform hypergraphs with controlled degree distribution. To produce a homogeneous degree distribution, we use the previous algorithm with the distribution $P(|e_j|) \sim \mathbbm{1}_{\{|e_j| = c\}}$, where $\mathbbm{1}_{\{|e_j| = c\}}$ is the indicator function which equals one if $|e_j| = c$ and zero otherwise. To produce a heterogeneous degree distribution, we use the algorithm proposed in~\cite{Jhun2019}, which generates uniform hypergraphs with PL distributions with $P(k) \sim k^{-\gamma}$ and $\gamma = 1 + \frac{1}{\nu}$. The algorithm associates each node $i$ with the probability $p_i = \frac{i^{-\nu}}{\zeta_N(\nu)}$, where $\zeta_N(\nu) = \sum_{j=1}^N j^{-\nu}$ and $0 < \nu < 1$. Next, for each hyperedge, we select $c$ nodes following the probabilities $p_i$. We again use the same brute force algorithm to ensure a single connected component, which may result in fluctuations similar to those observed in the random uncorrelated model.

\subsubsection{Numerical results}
\label{sec:numerical_res}

\begin{figure}[t!]
    \includegraphics[width=\linewidth]{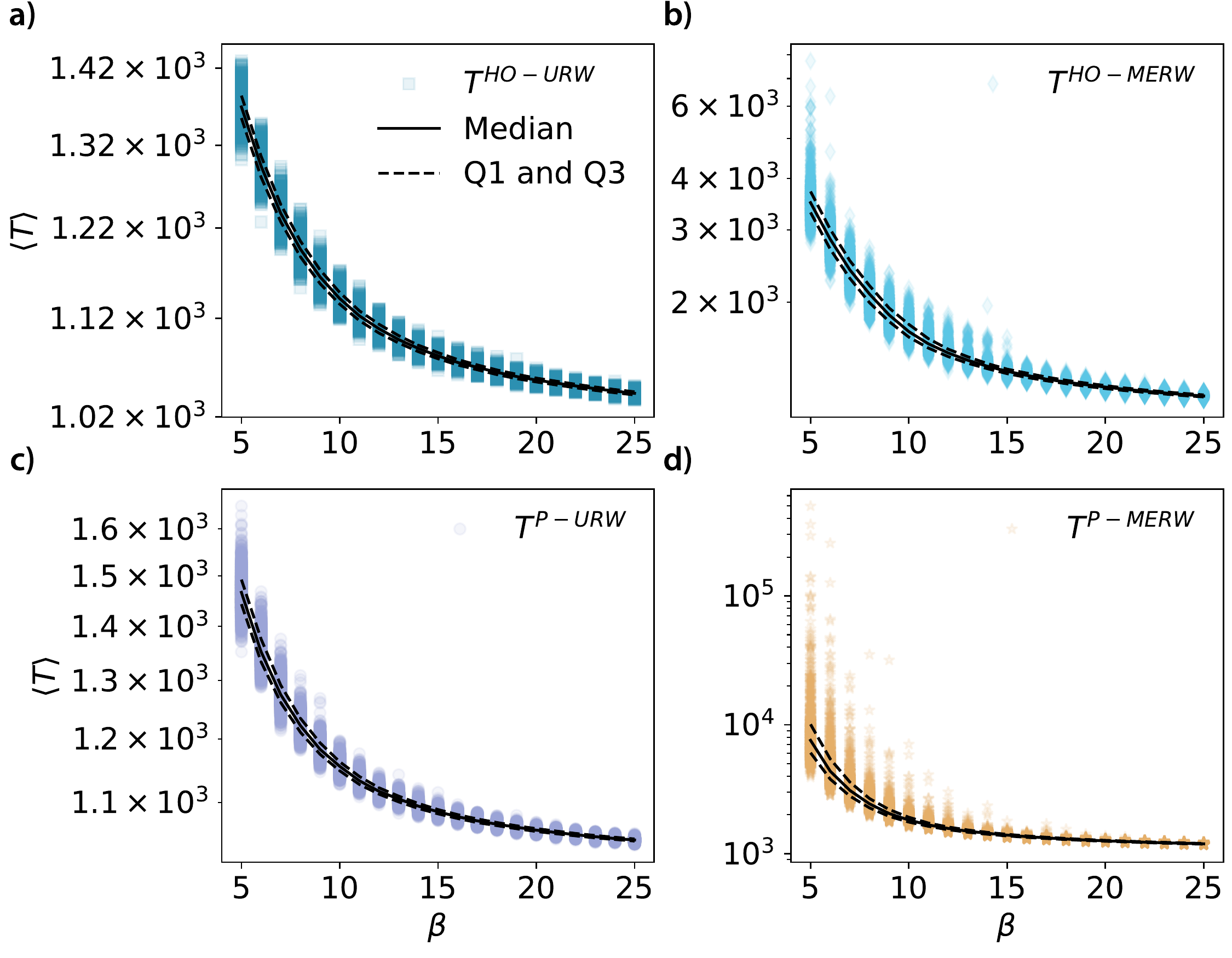}
    \caption{Mean hitting time on random hypergraphs following a Poisson distribution of cardinalities, $P(|e_j|) \sim \text{Poisson}(\beta)$, $N = 10^3$ and $M = 10^3$. For each parameter, we have $10^3$ independently generated hypergraphs, each one being a point in each panel. The median is represented by a black continuous line and the first and third quartiles are the dashed lines.}
    \label{fig:Poisson_Cardinalities}
\end{figure}

\begin{figure}[t!]
    \includegraphics[width=\linewidth]{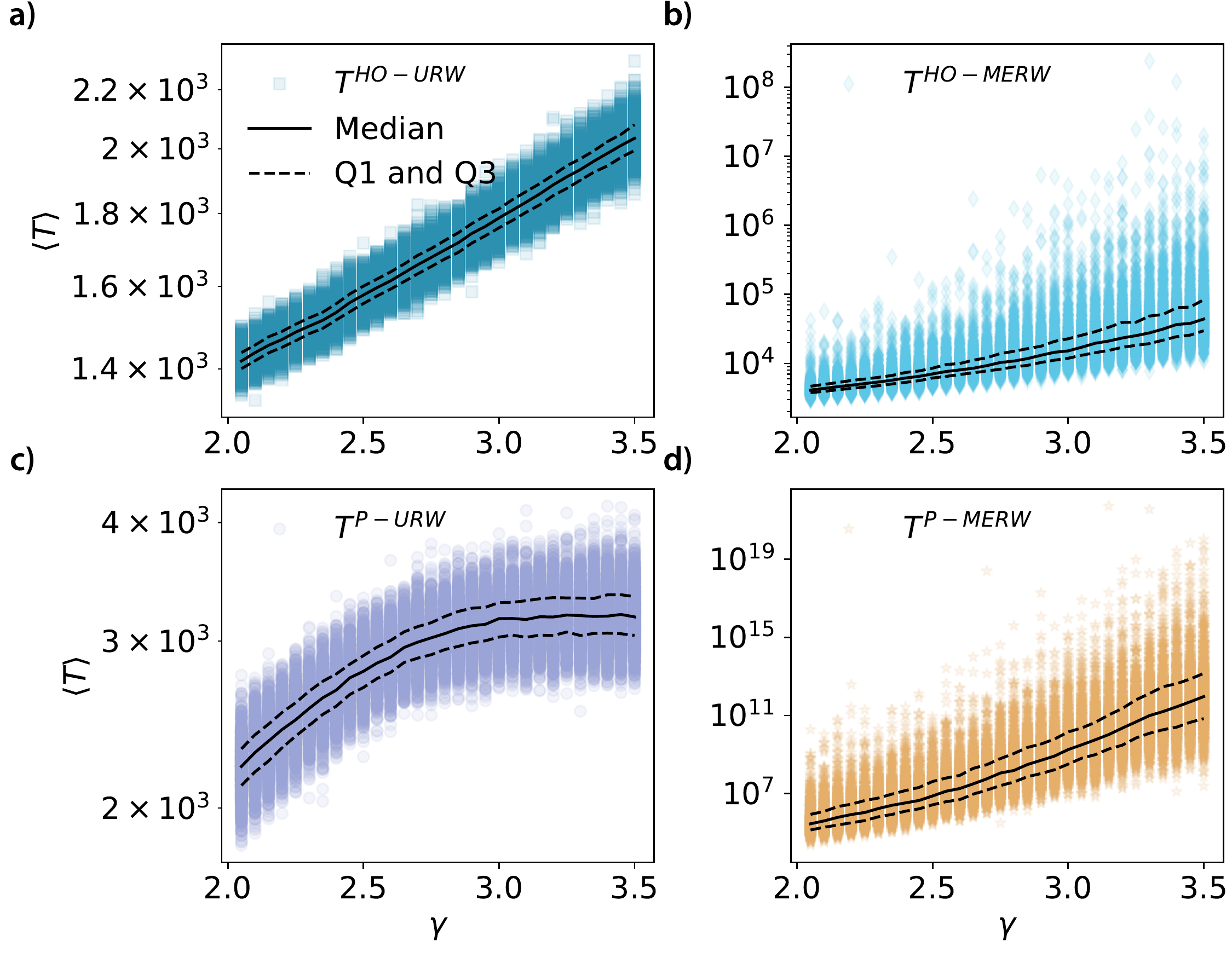}
    \caption{Mean hitting time on random hypergraphs following a power-law distribution of cardinalities, $P(|e_j|) \sim |e_j|^{-\gamma}$, $N = 10^3$ and $M = 10^3$. For each parameter, we have $10^3$ independently generated hypergraphs, each one being a point in each panel. The median is represented by a black continuous line and the first and third quartiles are the dashed lines.}
    \label{fig:PL_Cardinalities}
\end{figure}

\begin{figure}[t!]
    \includegraphics[width=\linewidth]{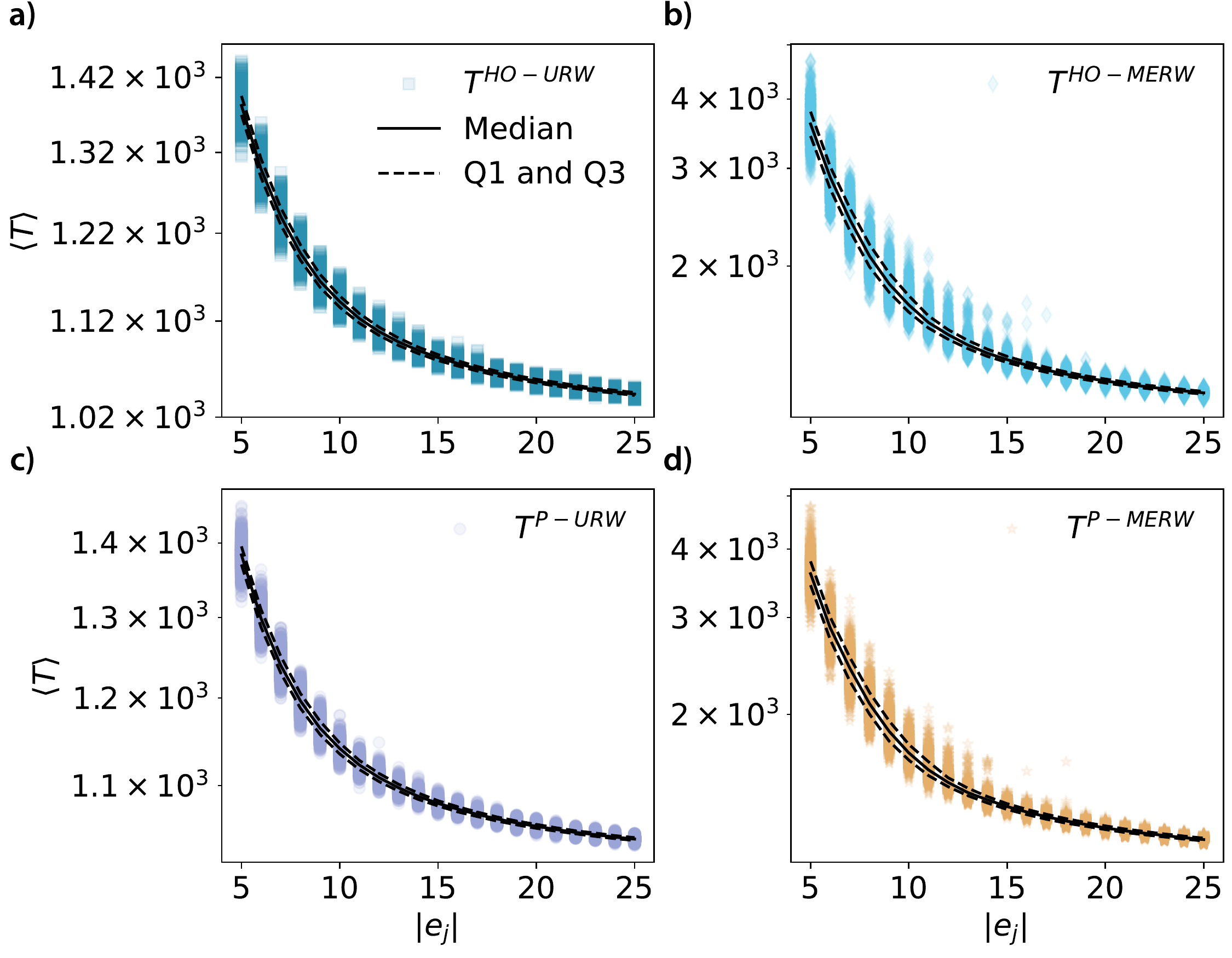}
    \caption{Mean hitting time on uniform homogeneous random hypergraphs, $N = 10^3$ and $M = 10^3$. For each parameter, we have $10^3$ independently generated hypergraphs, each one being a point in each panel. The median is represented by a black continuous line and the first and third quartiles are the dashed lines.}
    \label{fig:Uniform_Homogeneous}
\end{figure}

\begin{figure}[t!]
    \includegraphics[width=\linewidth]{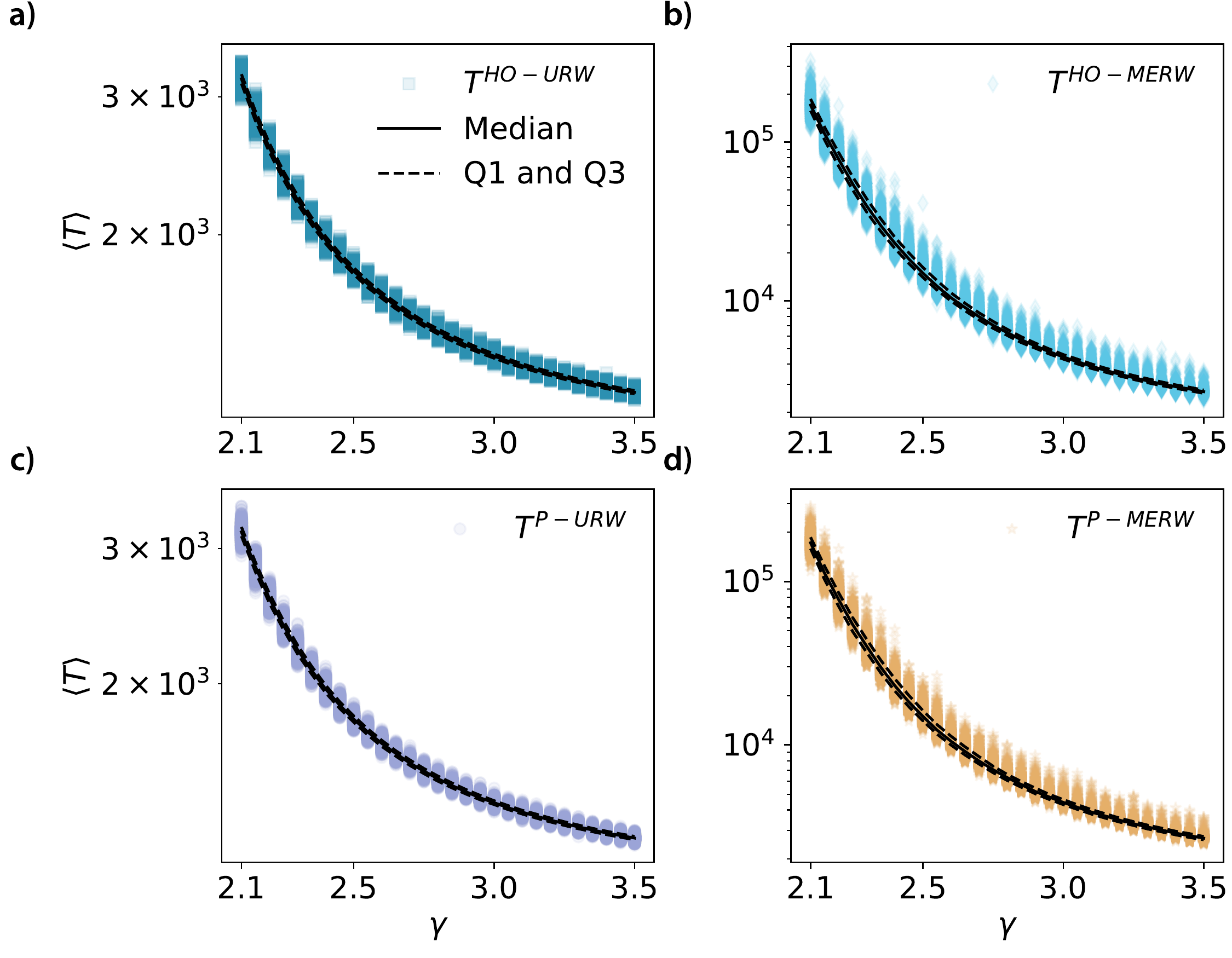}
    \caption{Mean hitting time on uniform random hypergraphs following a power-law degree distribution, $P(k)  \sim k^{-\gamma}$, $N = 10^3$, $|e_j| = 20$ and $M = 10^3$. For each parameter, we have $10^3$ independently generated hypergraphs, each one being a point in each panel. The median is represented by a black continuous line and the first and third quartiles are the dashed lines.}
    \label{fig:Uniform_PL}
\end{figure}

Fig.~\ref{fig:Poisson_Cardinalities} illustrates the mean hitting time, $\langle T \rangle$, for the four classes of random walks on different Poisson distributions of cardinality distributions and various values of $\beta$. Regardless of the type of random walk, $\langle T \rangle$ decreases as we increase $\beta$. We remark that $\langle T \rangle$ is lower bounded by $(N-1)$ as we need at least $(N-1)$ steps to visit all the nodes. Comparing Fig.\ref{fig:Poisson_Cardinalities} (a) and (c) with Fig.\ref{fig:Poisson_Cardinalities} (b) and (d), we observe that the URW has a smaller $\langle T \rangle$ than the MERW for both steps. Moreover, in general, $\langle T \rangle$ in the projected step is larger than the higher-order one.  As the structure is reasonably homogeneous, the URW is very similar for both steps. Although all curves show a similar trend, the MERW has a larger variance, particularly for small values of $\beta$. The extreme case is the P-MERW, where some structures have an average hitting time with a different order of magnitude, as shown in Fig.~\ref{fig:Poisson_Cardinalities} (d).

Complementary, in Fig.~\ref{fig:PL_Cardinalities}, we evaluate the impact of heterogeneity using a PL distribution of cardinalities. Except for the P-URW in Fig.~\ref{fig:PL_Cardinalities} (c), all the other cases present a mean hitting time that increases with $\gamma$. We also observe that the variance is relatively higher if compared to the Poisson case in Fig.~\ref{fig:Poisson_Cardinalities}. Similar comments as before also apply here, such as the projected step imposing a higher mean hitting time. Furthermore, we remark that the variance is more considerable for higher values of $\gamma$, which is particularly evident in the MERW cases. This phenomenon arises because increasing $\gamma$ with a fixed number of hyperedges $M$ results in a sparser projected hypergraph. Similar effects can be observed by reducing the parameter $\beta$ in Fig.~\ref{fig:Poisson_Cardinalities}, where the average size of each hyperedge decreases while the number of hyperedges is fixed, leading to a sparser projected hypergraph.

Next, we investigate the impact of degree distribution heterogeneity on uniform hypergraphs with fixed cardinalities. Figs. \ref{fig:Uniform_Homogeneous} and \ref{fig:Uniform_PL} show results for homogeneous and power-law degree distributions, respectively. We note that both types of steps in the URW produce the same output, due to the uniformity of the cardinalities. In this case, the counting and normalized adjacency matrices have the same spectral distributions, up to a scale. For the uniform homogeneous case in Fig. \ref{fig:Uniform_Homogeneous}, we observe that increasing the magnitude of the edges $|e_j|$ leads to a decrease in the mean hitting time, as also observed in the Poisson case, Fig.~\ref{fig:Poisson_Cardinalities}. In the PL case with $|e_j| = 20$ in Fig.~\ref{fig:Uniform_PL}, we find that, as we increase $\gamma$, the mean hitting time, $\langle T \rangle$, decreases. However, it's worth noting that the mean hitting times in the uniform homogeneous cases are considerably smaller, suggesting that degree distribution heterogeneity may also contribute to longer mean hitting times.

\subsection{Real hypergraphs}
\label{sec:real}

\begin{figure}[t!]
    \includegraphics[width=\linewidth]{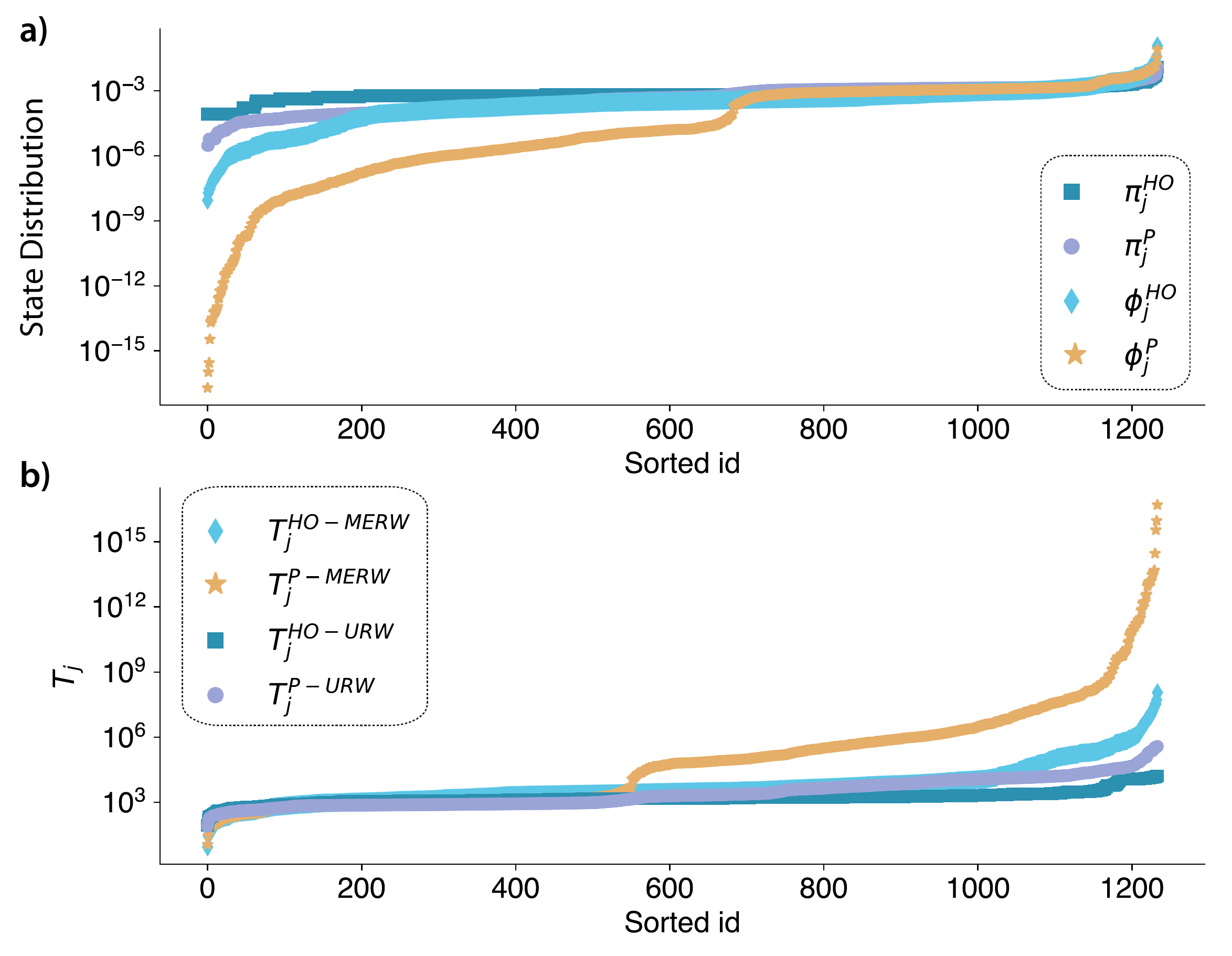}
    \caption{Analysis of the \emph{cat-edge-vegas-bars-reviews} real hypergraph. In (a) the stationary distribution for the four types of random walks, while in (b) Partial mean hitting time, $T_j$, for the same four types of random walks.}
    \label{fig:Real_0}
\end{figure}

\begin{figure}[t!]
    \includegraphics[width=\linewidth]{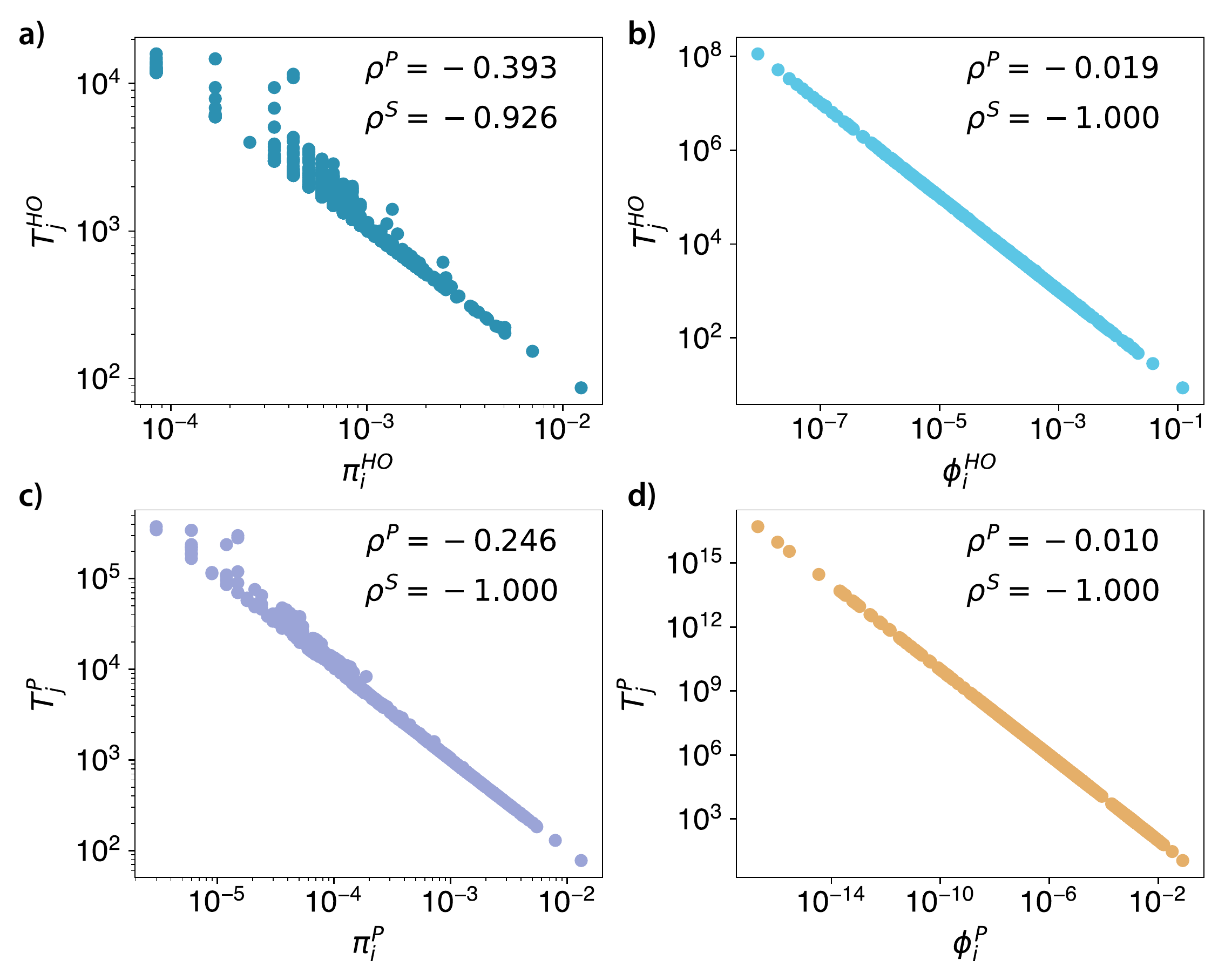}
    \caption{Comparative analysis of the different stationary distributions with their respective partial mean hitting time for the \emph{cat-edge-vegas-bars-reviews} real hypergraph. We report the Pearson, $\rho^P$, and Spearman, $\rho^S$, correlations in the upper right corner of each panel.}
    \label{fig:Real_2}
\end{figure}

\begin{figure}[t!]
    \includegraphics[width=\linewidth]{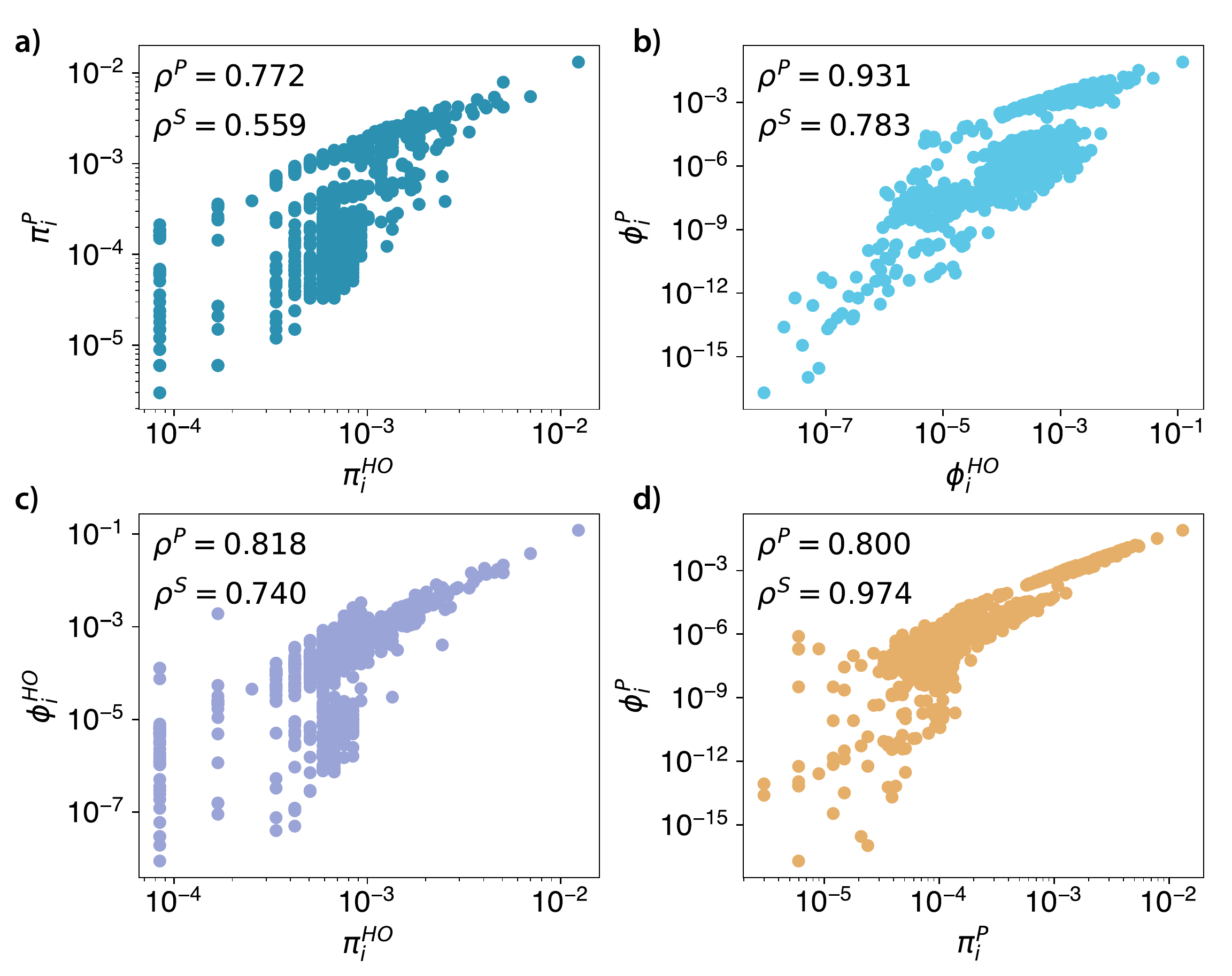}
    \caption{Comparative analysis of the different stationary distributions for the \emph{cat-edge-vegas-bars-reviews} real hypergraph.  We report the Pearson, $\rho^P$, and Spearman, $\rho^S$, correlations in the upper left corner of each panel.}
    \label{fig:Real_1}
\end{figure}

In this section, we present the analysis of the real hypergraph \emph{cat-edge-vegas-bars-reviews}~\cite{amburg2020fair}. Nodes are Yelp users, and hyperedges are users who reviewed an establishment of a particular category, which are different types of restaurants in Las Vegas, NV, United States. This data was collected for a month. We focus our analysis on this hypergraph, as it presented a very rich behavior, serving as an example of the different types of behavior introduced by correlations. Additional real hypergraphs are analyzed in Appendix~\ref{app:real}.

Figure~\ref{fig:Real_0} shows the stationary distribution and the partial mean hitting times in (a) and (b), respectively. In this figure, the x-axis is sorted independently for each type of random walks studied, for visualization purposes. We observe that the four types of random walks have very different behaviors. This outcome is especially evident for the P-MERW. In this case, we observe that in the MERW random walks, different nodes might play a very different role in the dynamics. Additionally, in Fig.~\ref{fig:Real_2}, we can verify that the stationary distributions and their respective partial mean hitting times are strongly correlated. However, this correlation is not linear as the Pearson correlation, $\rho^P$, is very small. However, the Spearman correlation is high, thus suggesting that there is a monotonous relationship. These effects can be observed in Fig~\ref{fig:Real_2}. From the analytical viewpoint, this is particularly evident for the MERW case, as in Eq.~\eqref{eq:Tj_ME}, $T_j$ is inversely proportional to the stationary distribution $\phi_i = \psi_i^2$. For the URW case, this argument is slightly more complex. A discussion about the behavior of URW hitting times and their scaling on networks can be found in Section 3.2.5 of~\cite{Masuda2017}.

Finally, in Fig.~\ref{fig:Real_1}, we compare the different classes of random walk in terms of their stationary distributions. In all the evaluated cases, the correlations are positive. Interestingly, comparing the two different steps with the same type of random walk, Fig.~\ref{fig:Real_1} (a) and (b), we observe that the Spearman correlation is considerably lower than the Pearson correlation. This result suggests that the relationship between different steps is non-trivial and that their rankings are not the same. Next, comparing the unbiased with the maximal entropy for each step, Fig.~\ref{fig:Real_1} (c) and (d), we observe a strong correlation. However, their relationship is not trivial.

\section{Analysis and discussion}
\label{sec:analysis}

At first glance, random walks on hypergraphs might seem like an abstract problem. However, similar to random walks in other contexts, this abstraction might provide insights into hypergraphs' structural organization. For instance, in~\cite{Carletti2020b}, random walks were used to detect community structures. As an application, the authors considered a hypergraph where nodes represent animals and hyperedges represent features, and they used random walks to group ``similar animals''  into communities~\cite{Carletti2020b}.  Another possible abstraction is the calculation of the probability and time necessary for a message to travel from node $i$ to node $j$. These examples support our argument that random walks are more general than their simplistic interpretation of a walker following a physical path in the hypergraph.

Here, we have shown that random walks can have different interpretations depending on the type of step adopted. We have considered what we call the projected and the higher-order steps. In the projected step, the hypergraph is effectively projected on a graph, where the walk takes place. This is exemplified by the toy hypergraph in Fig.~\ref{fig:Toy}, where we note that projected steps do not distinguish between a hyperedge with $|e_j| > 2$ and a clique with the $|e_j|$ nodes. In the higher-order step, the random walk can be interpreted as a sequence of two dependent processes, first a uniform choice of the next hyperedge, then choosing the next node. This construction allows for a distinction between the structures mentioned above. We remark that the projected and the higher-order steps are two possibilities, and many other processes can be defined following similar arguments. 
Indeed, although not explored here, in~\cite{Carletti2020b} the authors have formulated the random walk following the adjacency matrix
\begin{equation} 
    K_{ij}^{\tau} = 
    \begin{cases}
        \sum_{\alpha:e_{\alpha \in \mathcal{E}}} (\mathcal{B}_{\alpha \alpha} - 1)^{\tau}  \mathcal{I}_{i \alpha}  \mathcal{I}_{j \alpha}  \hspace{5mm} &\text{if} \hspace{2mm} i \neq j \\
        0 &\text{if} \hspace{2mm} i = j 
    \end{cases},
    \label{eq:K}
\end{equation}
where $\mathcal{B} = \mathcal{I}^T \mathcal{I}$, and $\tau$ is a real parameter. Note that, the element $\mathcal{B}_{ij}$ is the number of nodes in the intersection between $e_i$ and $e_j$, i.e., $|e_i \cap e_j|$. If $\tau = 0$ we recover the counting adjacency matrix and the projected step. On the other hand, if $\tau = -1$ we recover our normalized adjacency matrix and, thus the higher-order step. We can apply the same expressions introduced in Sec.~\ref{sec:RW} for the times and stationary distributions.

All these projections have in common that hyperedges of size $|e|$ are projected onto $|e|$-cliques. Another widely used and flexible alternative is to represent a hypergraph of $N$ vertices and $M$ hyperedges as a bipartite graph of $N+M$ nodes. The two classes of nodes in the bipartite graph are the vertices of the hypergraph and the hyperedges. 
The corresponding adjacency matrix is 
\begin{equation}\label{Abipartite}
    \mathbf{A}^{bipartite} =\begin{pmatrix}
\bigzero
& \mathcal{I} \\
\mathcal{I}^T 
& \bigzero
\end{pmatrix},
\end{equation}
where $\mathcal{I}$ is the hypergraph incidence matrix. It is worth noting that for this representation, a walker to go from a vertex $v_i$ to a vertex $v_j$ necessary has to move through a hyperedge that both vertices have in common, meaning that it will require 2 steps. The resulting walk is an alternating sequence of distinct pairs of vertices and hyperedges. Therefore, a step with the projected or higher-order step should be compared to two steps on the bipartite representation, as they both represent a movement from one vertex to another in the hypergraph. 

It's important to comment on the unbiased and maximal entropy random walk defined on this representation. We show that a non-trivial mapping exists between random walks defined on the bipartite representation of the hypergraph and random walks defined on top of the projections considered in Sec. \ref{Sec: representation}. Specifically, the unbiased random walk on the bipartite is equivalent to the unbiased random walk on the normalized adjacency projection. This equivalence was also pointed out in~\cite{mulas_random_2022} and is very intuitive. One can indeed notice that the weights appearing in Eq. \eqref{eq:adjacency_norm} represents a uniform choice between $|e|-1$ possible arrival nodes.
What is more interesting is the mapping between the MERW on the bipartite representation and the MERW with the projected step. To show it, we reproduce the derivations in~\cite{Burda2009}. We denote as $\gamma^{2t}_{v_{i_0},v_{i_t}}$ the trajectory of length $2t$ corresponding to the sequence $v_{i_0}e_{j_1},v_{i_1} \dots, e_{j_t}v_{i_t}$. The MERW is the random walk that maximizes the entropy of the set of sequences of length $2t$,

\begin{equation}\label{eq: Shannon entropy}
    S_t = - \sum_{v_{i_0}, e_{j_1}, \dots, v_{i_t}}  \Prob{v_{i_0}, e_{j_1}, \dots, v_{i_t}} \ln \Prob{v_{i_0}, e_{j_1}, \dots, v_{i_t}}.
\end{equation}

The probability of the sequence is $\Prob{v_{i_0}, e_{j_1}, \dots, v_{i_t}} = \pi_0 \Prob{\gamma^{2t}_{v_{i_0},v_{i_t}}}$ and $\pi_0$ is the probability of being in node $v_0$. This quantity is maximized when the sequence is chosen with uniform probability. We denote by $N_{2t}$ the number of all the possible sequences of length $2t$, then the entropy simplifies to
\begin{equation}
    S_t = \ln N_{2t}.
\end{equation}
The number of paths of length $2t$ between a pair of nodes $v_{i_0},v_{i_t}$ in the graph is simply given by $\left [ \left(\A ^{bipartite}\right)^{2t} \right ]_{i_0 i_t}$. Since we are working with non-lazy random walks, the above expression can be shown to be modified to $\left [ \left(\A ^{count}\right)^t \right ]_{i_0 i_t}$. As a consequence, the Shannon entropy in Eq. \eqref{eq: Shannon entropy} is maximized for

\begin{equation}
     S_t = \ln \sum_{v_{i_0}, v_{i_t}} \left [ \left(\A ^{count}\right)^t \right ]_{i_0 i_t} 
       \underset{t \rightarrow \infty}{\sim} t \ln \lambda ,  
\end{equation}
where $\lambda$ is the leading eigenvalue of $\A^{count}$. It can be shown that the correct choices for the transition probabilities of a non-lazy maximal entropy random walk on the bipartite projections are

\begin{align}
    P(v_i \rightarrow e) &= \frac{\mathcal{I}_{ie}}{\lambda} \frac{Z_i(e)}{\psi_i}, \\
    P(e \rightarrow v_k|v_i) &= \frac{ \psi_k \mathcal{I}_{ek}^T (1 - \delta_{ik} )}{Z_i(e)},\\
    Z_i(e) &= \sum_{k \neq i} \mathcal{I}_{ek}^T \psi_k.
\end{align}
$P(e \rightarrow v_k|v_i)$ is the probability to transition from hyperedge $e$ to node $v_k$ given that at the previous step the walker was in $v_i$ and $\psi_i$ is the $i$-component of the eigenvector corresponding to the leading eigenvalue $\lambda$.
This type of random walk maximizes the Shannon entropy in Eq. \eqref{eq: Shannon entropy} in the limit $t \rightarrow \infty$ and is equivalent to the maximal entropy random walk defined on top of the counting projection. Indeed, by considering the 2-step probability
\begin{equation}
    \begin{split}
        P(v_i \rightarrow v_j) &= \sum_e  P(v_i \rightarrow e)  P(e \rightarrow v_k|v_i) = \\
                                &= \sum_e   \frac{\mathcal{I}_{ie}}{\lambda} \frac{\mathcal{I}_{ej}^T (1 - \delta_{ij} ) \psi_j}{\psi_i} = \\
                                &= \frac{\A^{count}_{ij}}{\lambda} \frac{\psi_j}{\psi_i},
    \end{split}
\end{equation}
which is precisely the expression in Eq. \eqref{eq: MERW} with the projected step.

\section{Conclusions}
\label{sec:conclusion}

In this paper, we introduced maximal entropy random walks in hypergraphs, which to the best of our knowledge, has not been previously explored. Besides, we present a novel construction of random walks that complements the results in~\cite{Burda2009, Chitra2019, Carletti2020, Hayashi2020} by allowing for different types of steps. We explored the projected and higher-order steps to construct the unbiased and the maximal entropy random walks and characterize their stationary distribution, hitting times, partial and mean hitting times analytically. Our numerical experiments consider homogeneous and heterogeneous structures in terms of cardinality and degree distributions. We observe that, regardless of the type of random walk, increasing the average cardinality tends to decrease the average hitting time as seen in the numerical experiments reported in figures~\ref{fig:Poisson_Cardinalities} and~\ref{fig:Uniform_Homogeneous} for homogeneous cases. Furthermore, our experiments suggest that heterogeneity increases the mean hitting time for uniform hypergraphs with PL degree distribution. Notably, we observe that the hitting times for the projected step are typically larger than the higher-order step.

We also evaluate a real hypergraph with different types of correlations, emphasizing the complementary nature of the four classes of processes studied here, (P/HO)-(U/ME)RW, and providing different insights about the underlying structure. We discuss other possible types of steps found in the literature, particularly the walk on the bipartite representation of the hypergraph. We comment that a non-trivial mapping exists between random walks defined on the bipartite representation of the hypergraph and random walks defined on top of the projections considered in this paper. Specifically, the URW on the bipartite is equivalent to the URW on the normalized adjacency projection, while the MERW on the bipartite is equivalent to the MERW on the counting adjacency projection. Overall, our work contributes to a better understanding of random walks on hypergraphs and provides a versatile tool for analyzing complex systems in various domains.

Our results highlight the importance of the localization properties of the adjacency matrix and suggest that this feature might play a crucial role in other processes such as social contagion and information diffusion on hypergraphs. We hope that our findings will motivate further research with the potential to provide valuable insights for various applications, including the analysis of real higher-order systems and the development of novel methods in related fields such as artificial intelligence and behavioral sciences.

\begin{acknowledgments} 
Y.M was partially supported by the Government of Arag\'on, Spain and ``ERDF A way of making Europe'' through grant E36‐23R (FENOL), and by Ministerio de Ciencia e Innovaci\'on, Agencia Espa\~nola de Investigaci\'on (MCIN/AEI/10.13039/501100011033) Grant No. PID2020‐115800GB‐I00. We acknowledge the use of the computational resources of COSNET Lab at Institute BIFI, funded by Banco Santander (grant Santander‐UZ 2020/0274) and by the Government of Arag\'on (grant UZ-164255). The funders had no role in study design, data collection and analysis, the decision to publish, or the preparation of the manuscript. 
\end{acknowledgments}

\appendix

\begin{table*}[t!]\scalefont{0.95}
    \caption{Structural characterization of the real hypergraphs. The hypergraphs are characterized by the number of nodes, $N$, number of hyperedges, $M$, average cardinality, $\langle |e_j| \rangle$, standard deviation of the cardinalities $\text{std}(|e_j|)$, maximal cardinality, $\max (e_j)$. The normalized adjacency matrix metrics are the average, standard deviation minimum and maximum degree, $\langle k^{HO} \rangle$, $\text{std}(k^{HO})$, $k_{\min}^{HO}$ and $k_{\max}^{HO}$, while the respective metrics from the counting adjacency matrix are $\langle k^{P} \rangle$, $\text{std}(k^{P})$, $k_{\min}^{P}$ and $k_{\max}^{P}$. In the hypergraphs marked with a ``$*$'' we have repeated hyperedges. For more, see Appendix~\ref{sec:database}.}
    \begin{tabular}{|l|c|c|c|c|c|c|c|c|c|c|c|c|c|c|}
        \hline
        \hline
        Name & $N$ & $M$ & $\langle |e_j| \rangle$ & $\text{std}(|e_j|)$ & $\max (e_j)$  & $\langle k^{HO} \rangle$ & $\text{std}(k^{HO})$ & $k_{\min}^{HO}$ & $k_{\max}^{HO}$ & $\langle k^{P} \rangle$ & $\text{std}(k^{P})$ & $k_{\min}^{P}$ & $k_{\max}^{P}$\\
        \hline
        cat-edge-algebra-questions & 420 & 1267 & 6.519 & 6.579 & 107 & 19.664 & 34.091 & 1 & 375 & 239.076 & 352.769 & 1 & 3362 \\
        cat-edge-geometry-questions & 580 & 1193 & 10.465 & 15.647 & 230 & 21.526 & 36.264 & 1 & 260 & 707.334 & 1066.547 & 1 & 6711 \\
        cat-edge-vegas-bars-reviews & 1234 & 1194 & 9.937 & 13.817 & 73 & 9.615 & 7.371 & 1 & 147 & 270.665 & 295.724 & 1 & 4388 \\
        cat-edge-madison-restaurant-rev. & 565 & 601 & 7.656 & 7.281 & 43 & 8.143 & 7.217 & 1 & 59 & 110.588 & 104.189 & 2 & 716 \\
        cat-edge-music-blues-reviews & 1104 & 693 & 15.147 & 14.716 & 83 & 9.508 & 10.723 & 1 & 127 & 270.447 & 279.523 & 2 & 3393 \\
        phs-email-Enron & 4423 & 15653 & 4.119 & 4.458 & 25 & 14.576 & 101.395 & 1 & 4869 & 115.795 & 494.400 & 1 & 15471 \\
        phs-email-W3C & 13351 & 19351 & 2.219 & 0.953 & 25 & 3.217 & 24.751 & 1 & 958 & 5.237 & 31.534 & 1 & 1293 \\
        contact-high-school$^*$ & 327 & 172035 & 2.050 & 0.234 & 5 & 1078.648 & 816.639 & 7 & 4495 & 1161.639 & 883.960 & 7 & 4655 \\
        contact-primary-school$^*$ & 242 & 106879 & 2.096 & 0.310 & 5 & 925.612 & 446.772 & 125 & 2234 & 1056.744 & 530.606 & 131 & 2640 \\
        \hline
        \hline
    \end{tabular}
    \label{tab:database}
\end{table*}

\begin{figure*}[t!]
    \includegraphics[width=\linewidth]{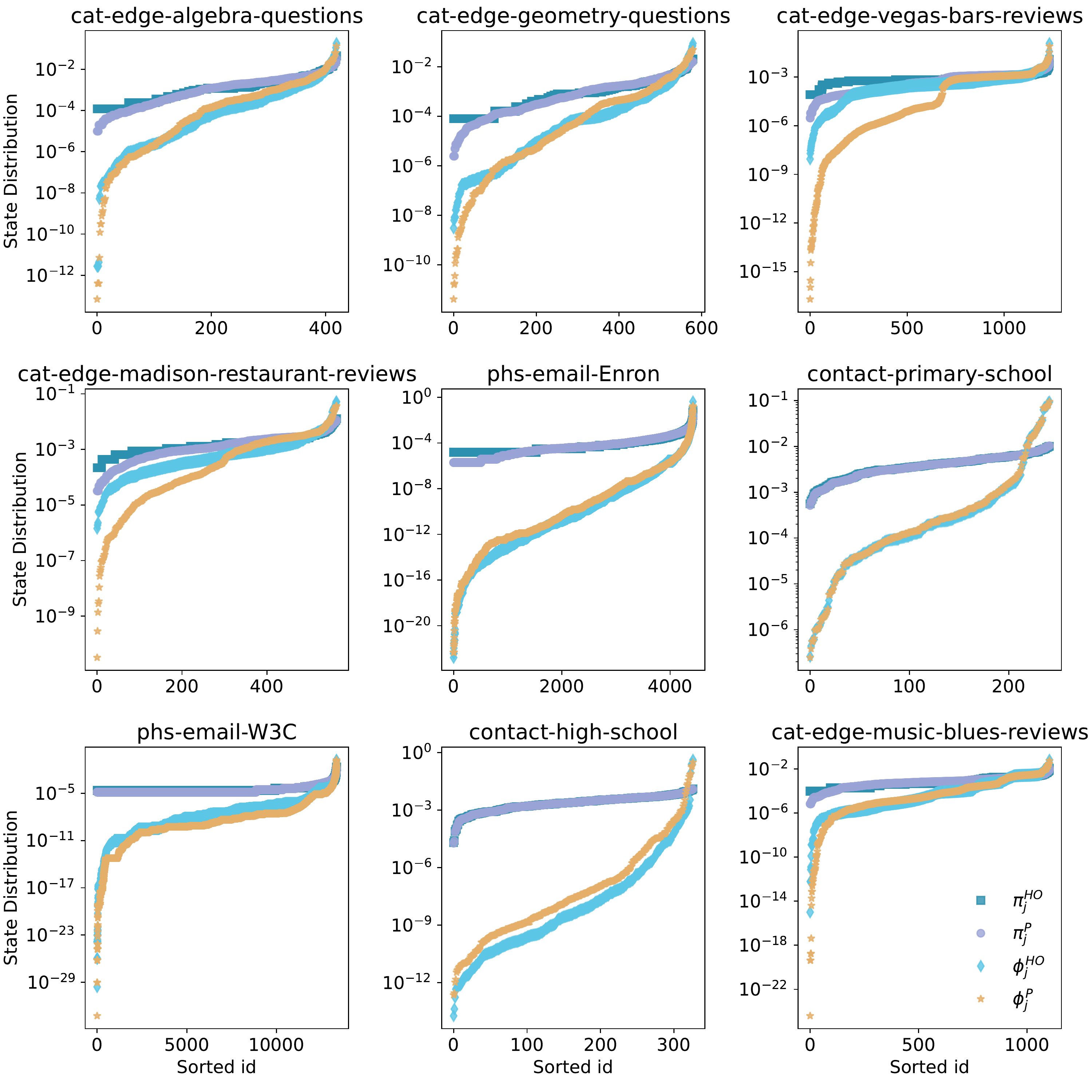}
    \caption{Stationary distribution for the four types of random walks studied for the database described in Section~\ref{sec:database}.}
    \label{fig:Densities_RD}
\end{figure*}

\begin{figure*}[t!]
    \includegraphics[width=\linewidth]{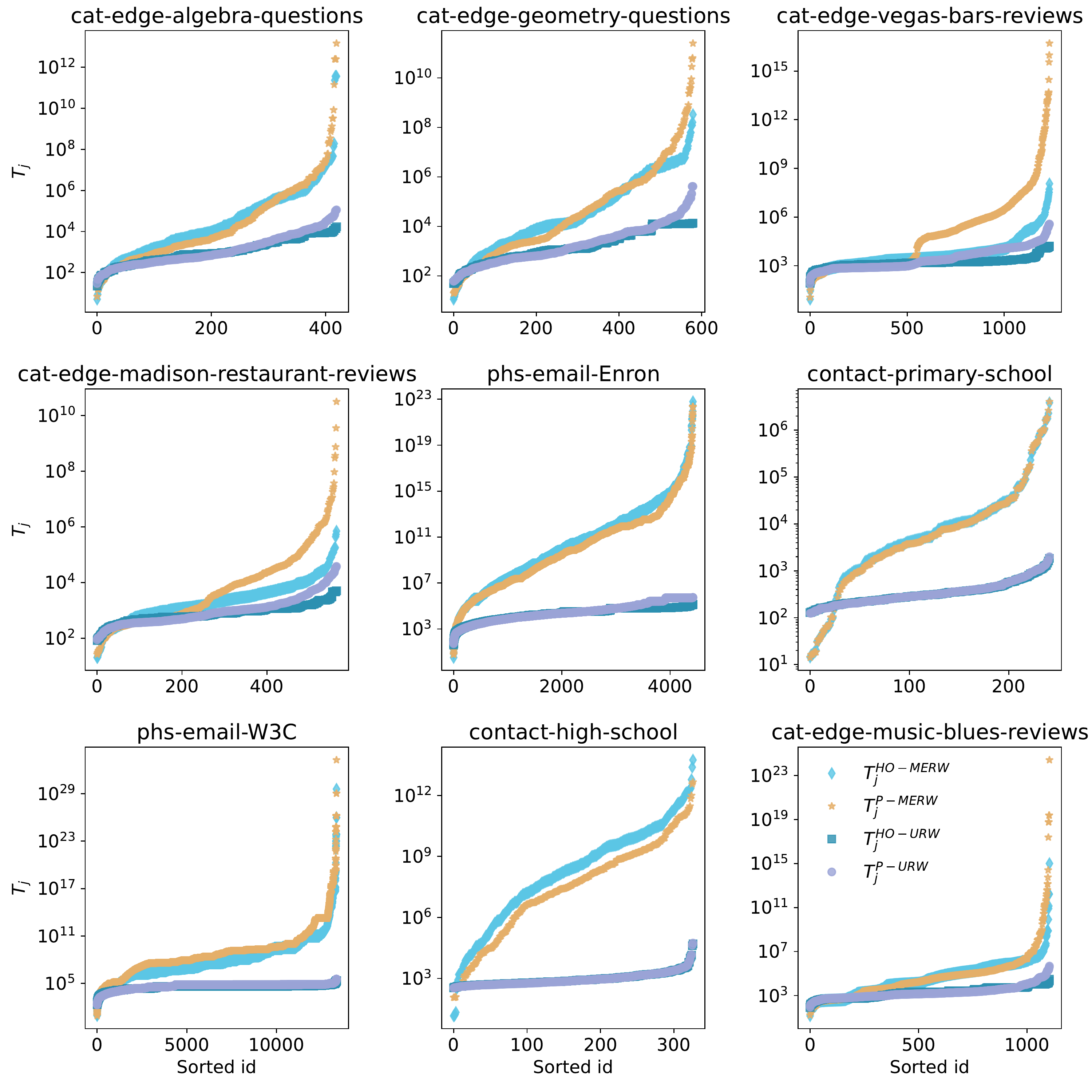}
    \caption{Partial mean hitting time, $T_j$, for the four types of random walks studied for the database described in Section~\ref{sec:database}.}
    \label{fig:Time_RD}
\end{figure*}

\section{Additional experiments: real hypergraphs}
\label{app:real}

In this section, we present additional experiments on real hypergraphs to complement the results of the main text. Here we focus on the stationary distributions and the partial mean hitting times. First, in the next section, we briefly describe the databases, while in Section~\ref{sec:additional}, we briefly compare their results.

\subsection{Databases}
\label{sec:database}

The database used here is hosted at \href{https://www.cs.cornell.edu/~arb/data/}{https://www.cs.cornell.edu/$\sim$arb/data/}. Additionally, in Table~\ref{tab:database}, we provide a brief structural characterization of these hypergraphs. We kept the dataset names on the repository to facilitate its identification, reproduction, and further studies. For more information about a specific hypergraph, please see the provided references. We also provide a short description of each hypergraph below:
\begin{itemize}
    \item \emph{cat-edge-algebra-questions}~\cite{amburg2020fair}: A hypergraph where nodes are users on MathOverflow and hyperedges are sets of users who answered a certain question category. This dataset was collected from different tags involving algebra, and it was derived from the Stack Exchange data dump;
    
    \item \emph{cat-edge-geometry-questions}~\cite{amburg2020fair}: A hypergraph where nodes are users on MathOverflow and hyperedges are sets of users who answered a particular question category. This dataset was collected from different tags involving geometry, and it was derived from the Stack Exchange data dump;
    
    \item \emph{cat-edge-vegas-bars-reviews}~\cite{amburg2020fair}: A hypergraph where the nodes are Yelp users and hyperedges are users who reviewed a bar of a particular category. This dataset is restricted to bars in Las Vegas, NV, and within a month's timeframe. The data was obtained from the Yelp Kaggle competition data;
    
    \item \emph{cat-edge-madison-restaurant-reviews}~\cite{amburg2020fair}: A hypergraph where the nodes are Yelp users and hyperedges are users who reviewed a restaurant of a particular category. This dataset is restricted to restaurants in Madison, WI, and within a month's timeframe. The data was obtained from the Yelp Kaggle competition data;
    
    \item \emph{cat-edge-music-blues-reviews}~\cite{ni2019justifying}: A hypergraph where nodes are Amazon reviewers and hyperedges are reviewers who reviewed a specific type of blues music within a month timeframe. The dataset was compiled from the product reviews collected by Jianmo Ni, Jiacheng Li, and Julian McAuley;
    
    \item \emph{phs-email-W3C}~\cite{Craswell-2005-TREC, Amburg-2019-planted-hitting-sets}: A hypergraph where nodes correspond to email addresses with a w3c.org domain and hyperedge consists of a set of email addresses, which have all appeared on the same email. This dataset was originally used on the analysis of core-fringe structures in~\cite{Amburg-2019-planted-hitting-sets};
    
    \item \emph{phs-email-Enron}~\cite{Amburg-2019-planted-hitting-sets}:  A hypergraph where nodes correspond to email addresses and hyperedges consist of sets of email addresses, which have all appeared on the same email. This dataset was originally used on the analysis of core-fringe structures in~\cite{Amburg-2019-planted-hitting-sets}, where core nodes correspond to email addresses of the individuals whose email inboxes were released as part of the investigation by the Federal Energy Regulatory Commission;
    
    \item \emph{contact-high-school}~\cite{Mastrandrea-2015-contact, Benson-2018-simplicial}: This dataset is a temporal sequence of timestamped hyperedges, which are composed of people. It is constructed from interactions recorded by sensors worn by people at a high school. The resolution of these sensors is 20 seconds;
    
    \item \emph{contact-primary-school}~\cite{Stehl-2011-contact, Benson-2018-simplicial}: This dataset is a temporal sequence of timestamped hyperedges, which are composed of people. It is constructed from interactions recorded by sensors worn by people at a primary school. The resolution of these sensors is 20 seconds;
\end{itemize}

We remark that \emph{contact-high-school} and \emph{contact-primary-school} have repeated hyperedges, thus the generated hypergraph is not simple. In the following section, we will use this version with repeated hyperedges to emphasize that our results also apply to this type of hypergraph. Thus, the interpretation of the random walk in these cases also slightly changes.

\subsection{Additional experiments}
\label{sec:additional}

Figure~\ref{fig:Densities_RD} shows the stationary distribution for the database discussed in Appendix~\ref{sec:database} and the four types of random walks studied. Note that, for visualization purposes, the x-axis is the sorted id, which is done independently for each curve, implying that the rankings might be different. Perhaps the most evident difference among the distributions is observed in the class of the process, URW or MERW, and later on the step's definition. Moreover, the MERW presents a higher variance of states, spanning orders of magnitude. Finally, this random walk, only in some cases, also presented ``lumps'', which are most visible for \emph{cat-edge-madison-restaurant-reviews} and \emph{cat-edge-vegas-bars-reviews} hypergraphs. These observations might suggest that these structures present some form of localization, imprisoning the walkers into ``entropic wells.'' 

Figure~\ref{fig:Time_RD} shows the partial mean hitting time for the database discussed in Appendix~\ref{sec:database} and the four types of random walks studied. Again, we highlight that the x-axis is independently sorted for each curve. For the partial mean hitting times, similar comments, as for Fig.~\ref{fig:Densities_RD}, also apply. As a particular observation for this measurement, we observe that $T_j$ is typically higher for the MERW. However, some nodes present a lower partial mean hitting time if compared to the URW. This result might suggest that these nodes play a notably different role, maybe serving as bridges.

\end{document}